\documentclass[aps,pra,superscriptaddress,floatfix,amsmath,twocolumn]{revtex4-2} 
\usepackage[utf8]{inputenc}
\usepackage{braket}
\usepackage{graphicx}
\usepackage[top=1in, bottom=1in, left=1in, right=1in]{geometry}
\usepackage{dsfont}
\usepackage{float}
\usepackage{amsmath}
\usepackage{xcolor}
\usepackage{natbib}
\usepackage{hyperref}
\usepackage{listingsutf8}
\definecolor{codegreen}{rgb}{0,0.6,0}
\definecolor{codegray}{rgb}{0.5,0.5,0.5}
\definecolor{codepurple}{rgb}{0.58,0,0.82}
\definecolor{backcolour}{rgb}{0.95, 0.95, 0.95}

\lstdefinestyle{codeblock}{
  backgroundcolor=\color{backcolour},
  commentstyle=\color{codegreen},
  keywordstyle=\color{blue},
  numberstyle=\tiny\color{codegray},
  stringstyle=\color{codepurple},
  basicstyle=\footnotesize,
  escapechar=\¢, 
  otherkeywords={with},
  breakatwhitespace=false,
  breaklines=true,
  captionpos=b,
  keepspaces=true,
  language=Python,
  numbers=right,
  numbersep=5pt,
  showspaces=false,
  showstringspaces=false,
  showtabs=false,
  tabsize=2,
  basicstyle=\ttfamily\footnotesize,
  inputencoding=utf8,
  upquote=true,
}
\begin{document}

\title{Quantum computing fidelity susceptibility using automatic differentiation}

\author{Olivia Di Matteo}
\affiliation{Department of Electrical and Computer Engineering, University of British Columbia, Vancouver, British Columbia V6T 1Z4, Canada}

\author{R. M. Woloshyn}
\affiliation{TRIUMF, Vancouver, British Columbia V6T 2A3, Canada}

\begin{abstract}
    Automatic differentiation is an invaluable feature of machine learning and quantum machine learning software libraries. In this work it is shown how quantum automatic differentiation can be used to solve the condensed-matter problem of computing fidelity susceptibility, a quantity whose value may be indicative of a phase transition in a system. Results are presented using simulations including hardware noise for small instances of the transverse-field Ising model, and a number of optimizations that can be applied are highlighted. Error mitigation (zero-noise extrapolation) is applied within the autodifferentiation framework to a number of gradient values required for computation of fidelity susceptibility and a related quantity, the second derivative of the energy. Such computations are found to be highly sensitive to the additional statistical noise incurred by the error mitigation method.
\end{abstract}

\maketitle

\section{Introduction}

Automatic differentiation is a technique for constructing a set of
instructions for computing the derivatives of a function defined by
a computer program. For classical computation it is implemented in software packages
such as Autograd \cite{autograd}, PyTorch \cite{pytorch}, TensorFlow \cite{tensorflow} and Jax \cite{jax2018github}. It is now a crucial component in 
machine learning libraries and has found diverse applications in 
physics and chemistry research. Some examples include analysis of
errors in lattice QCD simulations \cite{Ramos_2019}, Hamiltonian 
design in material science \cite{https://doi.org/10.48550/arxiv.2203.07157}, 
solution of quantum many-body problems using the numerical
renormalization group \cite{PhysRevResearch.4.013227} and
reconstruction of the neutron star equation of state 
\cite{https://doi.org/10.48550/arxiv.2209.08883}. In quantum
chemistry, automatic differentiation has been applied, for example,
to Hartree-Fock calculations \cite{Tamayo_Mendoza_2018,
https://doi.org/10.48550/arxiv.2203.04441}.

It is natural that automatic differentiation
is being replicated in quantum computing
frameworks. An increasing number of libraries, e.g., Qiskit \cite{Qiskit}, TensorFlow Quantum \cite{tfq}, and PennyLane \cite{bergholm2020pennylane}, provide automatic computation of gradients and higher derivatives 
mainly for optimization and machine learning applications. Quantum differentiable 
programming has also been implemented specifically for quantum chemistry 
in PennyLane \cite{arrazola2021differentiable}.

In this work we consider the use of quantum automatic differentiation
to solve a problem in condensed matter physics, namely, the calculation
of fidelity susceptibility \cite{PhysRevE.76.022101,PhysRevB.76.104420}. 
This is a quantity 
related to derivatives of a wave function overlap and has been used in the 
study of quantum phase transitions in a variety of condensed matter models.
The use of classical automatic differentiation for fidelity susceptibility was explored previously in \cite{xie2020automatic} as an application of an automatically differentiable eigensolver for real-valued matrices. 

Our focus here is not on the physics of this problem but primarily to provide
a novel end-to-end demonstration of how \emph{quantum} automatic differentiation can be used to solve it. The PennyLane library \cite{bergholm2020pennylane} is used and the 
cross-platform capability of this framework is illustrated by running 
PennyLane code on Qiskit simulators. Finally, this study exposes some of the 
issues that have to be dealt with in the NISQ era of quantum computation. 
The computation was designed to reduce sensitivity to noise as much as possible since gate error mitigation with automatic differentiation is a significant challenge.
We implement such mitigation and perform an analysis
of the complications that arise when doing so in a practical setting. To that end this work presents, to the best of our knowledge, the first study of how applying error mitigation affects the computation of energy derivatives with respect to Hamiltonian parameters. It is found to be very sensitive even for small systems and a limited amount of mitigation.

In Sect. \ref{subsec:quantum_gradients} we briefly review how quantum gradients can be computed automatically using parameter-shift rules. In Sect. \ref{sec:fs} the concepts of fidelity \cite{PhysRevE.74.031123} and fidelity susceptibility \cite{PhysRevE.76.022101,PhysRevB.76.104420} 
as used in condensed matter physics are introduced. They are a measure
of how a wave function changes when Hamiltonian parameters change.
Therefore, they can be useful in the study of quantum systems which
exist in different phases depending on interaction parameters. The
calculation of fidelity susceptibility is done in a variational framework
and its derivation in terms of derivatives with respect to variational
parameters is outlined. In order to complete the calculation, one
needs to know how variational parameters change with respect to changes in
Hamiltonian parameters. For this, derivatives of energy expectation values are
required. This is also discussed in Sect. \ref{sec:tfim}. 

The model used in this study is the one-dimensional transverse field
Ising model. It is the simplest model which displays a quantum phase
transition and is discussed, for example, in \cite{Suzuki2013}. The 
model is exactly solvable by transforming it to a model of free fermions 
\cite{PFEUTY197079}
but that would defeat our purpose, so we work with the spin degrees of
freedom directly. In simulations that include a hardware noise model
only small systems (4 or 6 spins) are feasible. The techniques
used to simplify the calculation and reduce noise sensitivity are
discussed in Sect.~\ref{sec:tfim} and results of simulations are 
presented in Sect.~\ref{sec:simres}. A demonstration of automatically-differentiable error 
mitigation is presented in Sect.~\ref{sec:error_mitigation}. Alternate algorithms for calculating wave function overlaps are discussed in Appendix A, and supplementary information regarding error-mitigated gradients and additional results are presented in Appendices B and C respectively. 

\section{Background}

\subsection{Quantum gradients and automatic differentiation}
\label{subsec:quantum_gradients}

Let $U(\theta)$ be a parametrized (variational) quantum circuit with $\theta = \{\theta_i\}$ a set of real-valued parameters. Consider a quantum computation that begins with a set of qubits in the all-zeros state $\ket{0\cdots0}$, executes $U(\theta)$, and measures the expectation value of a Hamiltonian $H$. The analytical expression of the expectation value is a function of $\theta$,
\begin{equation}
E(\theta) = \langle H \rangle = \langle 0 \cdots 0 | U^\dagger (\theta)  H U(\theta) |0\cdots 0 \rangle.
\end{equation}

\noindent This function can be differentiated with respect to the variational parameters. A common means of achieving this is using \emph{parameter-shift rules}. 

Suppose we would like to compute the gradient of $E(\theta)$ with respect to a particular $\theta_i$. If $\theta_i$ appears in the circuit as the parameter of a single-qubit rotation, it is by now well-known that the gradient can be obtained by running the circuit at two different values of the parameter and combining the results according to the shift rule \cite{mitarai2018quantum, li2017hybrid, wierichs2022general, schuld2019evaluating, mari2021estimating}:
\begin{multline}
\frac{\partial E(\theta)}{\partial \theta_i} = \frac{1}{2} [ E(\ldots, \theta_i + \pi / 2, \ldots) - \\ E(\ldots, \theta_i - \pi/2, \ldots) ], \label{eq:param_shift}
\end{multline}
\noindent where it is implicit that all parameters other than $\theta_i$ are held constant during this process.

Such rules are advantageous as they allow for computation of gradients in both software and on hardware without modifying the actual structure of the circuit, as only the input parameters change. For more sophisticated unitary operations, generalized parameter-shift rules can be used \cite{wierichs2022general}, which may require evaluating the circuit at four, or even more shifted values, depending on the operation.

Parameter-shift rules provide a straightforward recipe for computing quantum gradients; they can be further differentiated in order to compute similar rules for higher-order derivatives, such as Hessians \cite{mari2021estimating}. The availability of these rules thus allows gradient computation to be incorporated into automatic differentiation frameworks, wherein parameters can be modified through classical processing (e.g., multiplication by a constant or by another parameter), and the chain rule can be applied. Results of fully-differentiable quantum computations can also be incorporated into more extensive hybrid classical-quantum computations.

\subsection{Fidelity susceptibility}\label{sec:fs}

Phase transitions are a ubiquitous feature of many-body systems. The
long-standing paradigm of describing phase transitions in terms of a 
local order parameter associated with symmetry breaking \cite{itzykson_drouffe_1989}
is not adequate for describing many of the systems studied in contemporary 
condensed matter physics \cite{sachdev_2011,Bernevig+2013}. Many alternative
ways of describing phase transitions have been proposed including
fidelity \cite{PhysRevE.74.031123} and fidelity susceptibility \cite{PhysRevE.76.022101,PhysRevB.76.104420}.

Consider a Hamiltonian with a single variable parameter
\begin{equation}
H(r)=H_{0}+rH_{1}. \label{eq:basic_ham}
\end{equation}
 When $r$ is varied the system may be in different phases. The fidelity,
as defined by Chen \emph{et al.} \cite{PhysRevA.77.032111} is
\begin{equation}
F(r,\delta)=\left|\left\langle \psi_{0}(r)|\psi_{0}(r+\delta)\right\rangle \right|,
\end{equation}
 where $\psi_{0}(r)$ and $\psi_{0}(r+\delta)$ are eigenstates (only ground
states are considered here) at two different values of the Hamiltonian
parameter. The susceptibility is the second derivative of this quantity
w.r.t. $\delta$ evaluated at $\delta$ = 0 \cite{PhysRevA.77.032111}
\begin{equation}\label{eq:sus}
\mathcal{S}(r)=\partial_{\delta}^{2}F(r,\delta)|_{\delta=0}.
\end{equation}
By expanding $\psi(r+\delta)$
in eigenstates of $H(r)$ one can derive a kind of spectral representation
of the fidelity susceptibility which involves the sum over a complete
set of states. For the ground state, the expression for the fidelity
susceptibility is \cite{PhysRevA.77.032111}
\begin{equation}
\mathcal{S}(r)=\sum_{n\neq0}\frac{\left|\left\langle \psi_{n}(r)\left|H_{1}\right|\psi_{0}(r)\right\rangle \right|^{2}}{\left|E_{0}(r)-E_{n}(r)\right|^{2}}.
\end{equation}
This, in principle, requires knowledge of the complete spectrum. However,
the use of automatic differentiation enables calculation of
the fidelity susceptibility using only the ground state wave function.

We suppose that the ground state wave function $\psi$ (now written dropping the qualifier 0) has been obtained by
a variational calculation. Let $\theta = \{\theta_{i}\}$ denote the variational
parameters which at the variational minimum are implicit functions
of $r.$ The fidelity susceptibility (\ref{eq:sus}) can be expressed as 
\begin{equation}\label{eq:d2over}
\mathcal{S}(r)=\left|\partial_{r}^{2}\left\langle \psi|\psi(\theta(r))\right\rangle \right|
\end{equation}
where it is understood that differentiation is applied only to the
right side of the overlap. By the chain rule
\begin{equation}
\partial_{r}\left\langle \psi|\psi(\theta(r))\right\rangle =\sum_{i}\frac{\partial\left\langle \psi|\psi(\theta(r))\right\rangle }{\partial\theta_{i}}\frac{\partial\theta_{i}}{\partial r},
\end{equation}
 and 
\begin{multline}\label{eq:d2chain}
\partial_{r}^{2}\left\langle \psi|\psi(\theta(r))\right\rangle =\sum_{i,j}\frac{\partial^{2}\left\langle \psi|\psi(\theta(r))\right\rangle }{\partial\theta_{j}\partial\theta_{i}}\frac{\partial\theta_{i}}{\partial r}\frac{\partial\theta_{j}}{\partial r}+\\\sum_{i}\frac{\partial\left\langle \psi|\psi(\theta(r))\right\rangle }{\partial\theta_{i}}\frac{\partial^{2}\theta_{i}}{\partial r^{2}}.
\end{multline}
With a parametrization of the variational wave function and a circuit
to compute the overlap, the derivatives of
the overlap w.r.t. the variational parameters can be computed in a straightforward manner through parameter-shift rules. Obtaining derivatives
of the variational parameters w.r.t. the Hamiltonian parameter $r$
is more involved.

For the transverse field Ising model (considered in this work) or other Hamiltonians for which
the expansion of the ground state wave function involves real coefficients, the gradient of the overlap w.r.t. variational parameters
will vanish so the second term in (\ref{eq:d2chain}) will not contribute. 
If the expansion coefficients are complex this may not be the case.

When the gradient 
\begin{equation}\label{eq:d1ovr}
 \frac{\partial\left\langle \psi|\psi(\theta(r))\right\rangle }{\partial\theta_{i}}
 \end{equation}
vanishes the fidelity susceptibility
can be obtained by computing the overlap Hessian and the gradients of the variational parameters. The Hessian can be calculated directly
or  alternatively, by considering $\left|\left\langle \psi|\psi(\theta(r))\right\rangle \right|^{2}$.
When (\ref{eq:d1ovr}) vanishes, then
\begin{equation}\label{eq:d2ovsqr}
\frac{\partial^{2}\left|\left\langle \psi|\psi(\theta(r))\right\rangle \right|^{2}}{\partial\theta_{j}\partial\theta_{i}}=2\frac{\partial^{2}\left\langle \psi|\psi(\theta(r))\right\rangle }{\partial\theta_{j}\partial\theta_{i}}
\end{equation}
is another means by which to compute the overlap Hessian. The choice
of algorithm for computing the Hessian is discussed 
in Sect. \ref{sec:simres}  and in Appendix A.

In order to complete the calculation of the fidelity susceptibility
the quantities $\frac{\partial\theta_{i}}{\partial r}$ are needed.
To get these, derivatives of the Hamiltonian expectation value are
used. See the article by Pulay \cite{pulay1987} for a systematic discussion.
The energy is
\begin{equation}
E=\left\langle \psi(\theta(r))\left|H\right|\psi(\theta(r))\right\rangle 
\end{equation}
 and the condition for the variational minimum is
\begin{equation}\label{eq:var}
\frac{\partial E}{\partial\theta_{i}}=\frac{\partial\left\langle \psi\left|H\right|\psi\right\rangle }{\partial\theta_{i}}=0
\end{equation}
 for all $i$. The first energy derivative is
\begin{eqnarray}
\frac{\partial E}{\partial r} & = & \left\langle \psi\left|\frac{\partial H}{\partial r}\right|\psi\right\rangle +\sum_{i}\frac{\partial\left\langle \psi\left|H\right|\psi\right\rangle }{\partial\theta_{i}}\frac{\partial\theta_{i}}{\partial r}\\
 & = & \left\langle \psi\left|H_{1}\right|\psi\right\rangle 
\end{eqnarray}
at the variational minimum. The second energy derivative is 
\begin{equation}\label{eq:d2E}
\frac{\partial^{2}E}{\partial r^{2}}=\sum_{i}\frac{\partial\left\langle \psi\left|H_{1}\right|\psi\right\rangle }{\partial\theta_{i}}\frac{\partial\theta_{i}}{\partial r},
\end{equation}
for a Hamiltonian which is linear in the variable parameter. 

To get the derivatives of the variational parameters w.r.t. $r$, take
the derivative of the variational condition (\ref{eq:var}). The first 
order response is 
\begin{multline}
\frac{\partial}{\partial r}\left(\frac{\partial\left\langle \psi\left|H\right|\psi\right\rangle }{\partial\theta_{i}}\right) = \frac{\partial\left\langle \psi\left|H_{1}\right|\psi\right\rangle }{\partial\theta_{i}}\\+\sum_{j}\frac{\partial^{2}\left\langle \psi\left|H\right|\psi\right\rangle }{\partial\theta_{i}\partial\theta_{j}}\frac{\partial\theta_{j}}{\partial r} = 0 \label{eq:first_order_response}.
\end{multline}
 This gives a set of equations that can be solved to get the $\frac{\partial\theta_i}{\partial r}.$
This is sufficient for the calculation of the first and second energy
derivatives as discussed in \cite{arrazola2021differentiable, mitarai2020theory} and for
the fidelity susceptibility calculated here. If second derivatives of the 
variational parameters are required they can be obtained by taking another 
derivative of the variational condition \cite{pulay1987}
\begin{multline}
\frac{\partial^{2}}{\partial r^{2}}\left(\frac{\partial\left\langle \psi\left|H\right|\psi\right\rangle }{\partial\theta_{i}}\right) = \sum_{j}\frac{\partial^{2}\left\langle \psi\left|H_{1}\right|\psi\right\rangle }{\partial\theta_{i}\partial\theta_{j}}\frac{\partial\theta_{j}}{\partial r} \\ + \sum_{j,k}\frac{\partial^{3}\left\langle \psi\left|H\right|\psi\right\rangle }{\partial\theta_{i}\partial\theta_{j}\partial\theta_{k}}\frac{\partial\theta_{j}}{\partial r}\frac{\partial\vartheta_{k}}{\partial r}\nonumber \\ +  \sum_{j}\frac{\partial^{2}\left\langle \psi\left|H\right|\psi\right\rangle }{\partial\theta_{i}\partial\theta_{j}}\frac{\partial^{2}\theta_{j}}{\partial r^{2}} \nonumber 
 = 0.
\end{multline}

\noindent In summary, to compute the fidelity susceptibility (second derivative of the
overlap) and the second derivative of energy, one needs to compute the following quantum gradients:
\begin{enumerate}
 \item The Hessian of the overlap w.r.t. the variational parameters
 \item The Hessian of the expectation value of the full Hamiltonian w.r.t. the
   variational parameters
 \item The gradient of the expectation value of the $r$-dependent part of the
   Hamiltonian ($H_1$) w.r.t. the variational parameters
 \end{enumerate}

\begin{figure}[tb]
\centering
\includegraphics[width=\columnwidth]{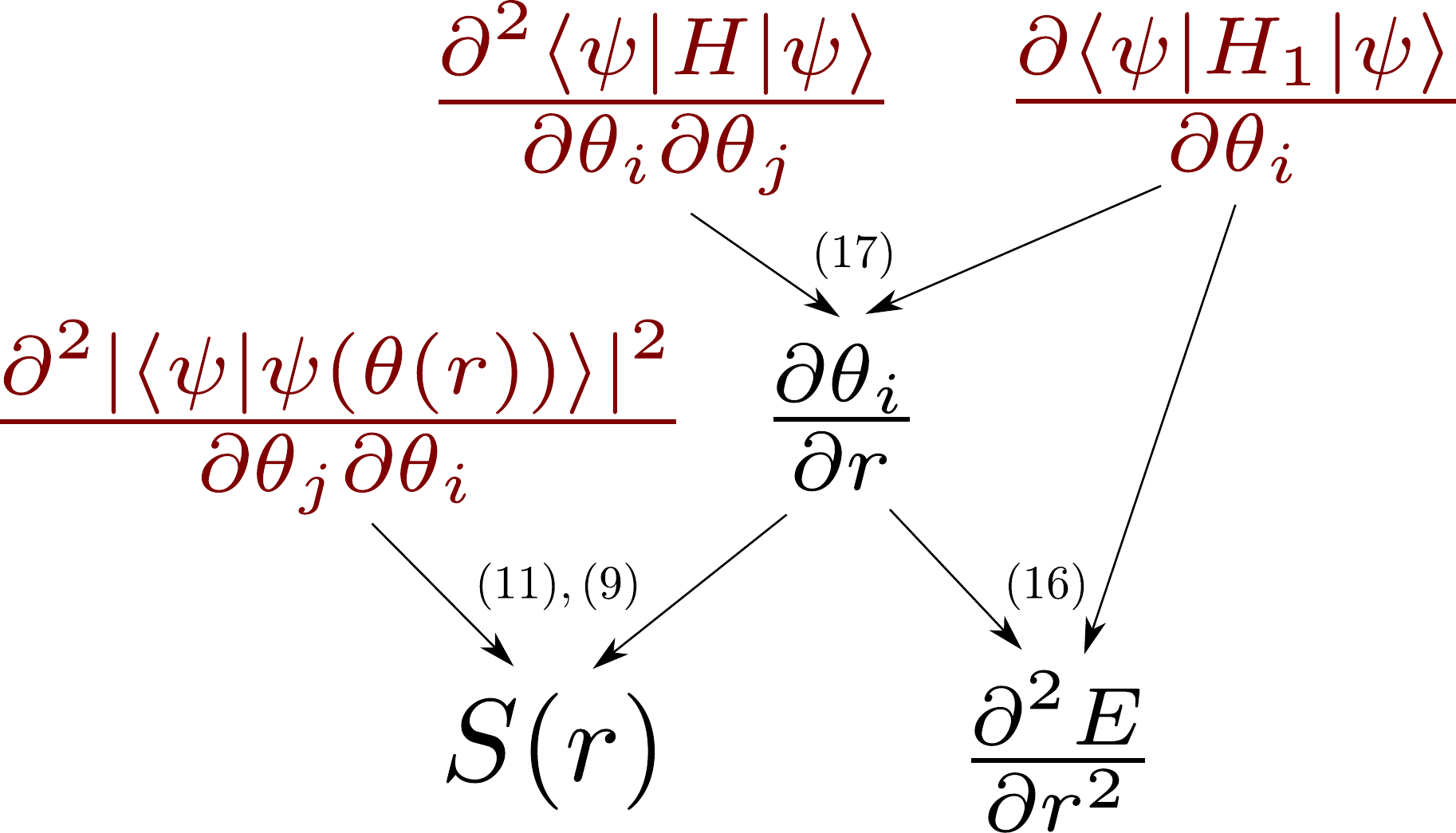}
\caption{Graphical depiction of how the quantities of interest depend on quantum gradients. The three gradients in red are obtained through quantum computation using parameter-shift rules, while the derivatives of the variational parameters with respect to $r$ are obtained by solving the response equations. Numbers in parentheses correspond to equations in the main text which are used to obtain the desired quantity.} 
\label{fig:flowchart}
\end{figure}

\noindent The interplay between all these gradients and how they feed into our quantities of interest is displayed graphically in Fig. \ref{fig:flowchart}. In Sect. \ref{sec:simres}, we will demonstrate how the necessary gradients can be computed using quantum automatic differentiation software.

\section{Fidelity susceptibility and the Transverse Field Ising Model}\label{sec:tfim}

This section outlines the analytical description of the quantum computation that will be performed. For the one-dimensional transverse field Ising model the Hamiltonian
\begin{equation}
H(r)=-\sum_{i=0}^{L-1}\left(\sigma_{i}^{x}\sigma_{i+1}^{x}+r\sigma_{i}^{z}\right)
\end{equation}
will be used with periodic boundary conditions, so there is a translation
invariance. Furthermore, the Hamiltonian is invariant under a transformation
by $\sigma^{z}$ at every site, that is, 
\begin{equation}\label{eq:HP}
\left[H(r),\mathcal{P}\right]=0,
\end{equation}
 where 
\begin{equation}
\mathcal{P}=\bigotimes_{i} \sigma_{i}^{z}.
\end{equation}

Consider a one-dimensional spin chain with an even number $L$ of sites.
The total number of possible spin configurations is $2^{L}.$ However,
due to (\ref{eq:HP}), only half of the possible spin configurations, i.e.,
$2^{(L-1)}$ contribute to the eigenstate of $H.$ For the ground state,
the contributing spin configurations are those where both the number
of up and down spins is even. Due to translation invariance many spin
configurations will have the same coefficient. These features are
used to simplify the quantum computation.

As an illustration of how the computation can be simplified consider
the case of 4 spins. 
The Hilbert space is spanned by the 16 possible spin configurations.
However, using symmetry arguments one can conclude that the 
ground state wave function lies in a much lower-dimensional subspace.
Due to the invariance (\ref{eq:HP}) of the Hamiltonian
only 8 of the spin configurations appear in the ground state wave function. Then translation invariance,
due to periodic boundary conditions,  allows these 8 spin
configurations to be grouped into 4 composite basis states:
\begin{eqnarray*}
 & \left|0000\right\rangle, \\
 & \frac{1}{2}\left(\left|0011\right\rangle +\left|1001\right\rangle +\left|1100\right\rangle +\left|0110\right\rangle \right),\\
 & \frac{1}{\sqrt{2}}\left(\left|0101\right\rangle +\left|1010\right\rangle \right),\\
 & \left|1111\right\rangle .
\end{eqnarray*}
Thus the wave function can be expressed as a superposition of only 4 independent states so only two qubits are required to encode it.
 Calculating the matrix elements of the full Hamiltonian using the
above states gives a reduced $(4\times4)$ Hamiltonian matrix
\begin{equation}\label{eq:ham4}
\left(\begin{array}{cccc}
-4r & 0 & -2 & 0\\
0 & 0 & -2\sqrt{2} & 0\\
-2 & -2\sqrt{2} & 0 & -2\\
0 & 0 & -2 & 4r
\end{array}\right).
\end{equation}
 It can be verified by exact diagonalization that the ground state energy
and fidelity susceptibility using the reduced Hamiltonian are identical
to the results obtained with the full Hamiltonian. 

\begin{figure}[tb]
\centering
\includegraphics[scale=0.35,trim={6cm 4cm 6cm 4cm},clip=true,angle=-90]{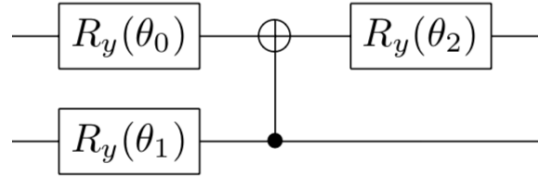}
\caption{Circuit for the 4-spin variational wave function encoded in two qubits. } 
\label{fourspin}
\end{figure}

\begin{figure}[t]
\centering
\includegraphics[width=\columnwidth]{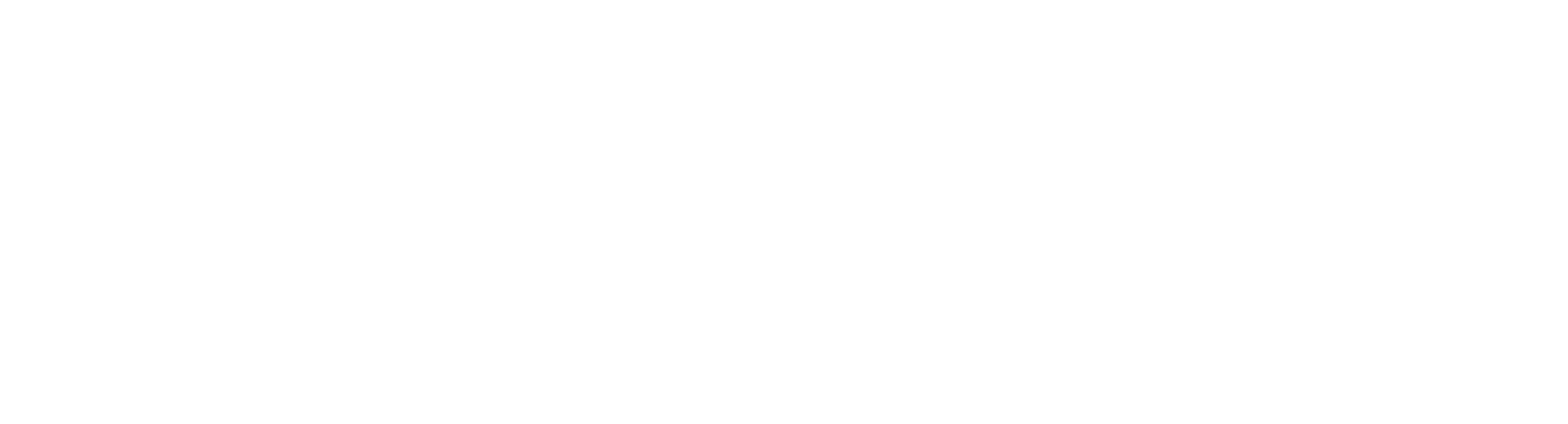}
\caption{Optimized circuit for the 6-spin variational wave function encoded in three qubits. } 
\label{sixspin}
\end{figure}

The quantum computation using the reduced Hamiltonian can be done
using two qubits with a variational wave function depending on three
parameters. The circuit for the unitary operator used to instantiate the 4-spin 
variational wave function is shown in Fig. \ref{fourspin}. 
By appropriate choice of parameters, the operator encoded in 
Fig. \ref{fourspin} can produce any possible superposition of 
four states with real coefficients. Therefore, it is a suitable
ansatz for the variational ground state of the Hamiltonian (\ref{eq:ham4}).
In terms of Pauli operators 
the Hamiltonian (\ref{eq:ham4}) acting on two qubits is
\begin{multline}
-X(0)-X(1)-X(0)Z(1)+Z(0)X(1) \\ -\sqrt{2}(X(0)X(1)+Y(0)Y(1))-2r(Z(0)+Z(1)), \nonumber
\end{multline}

\noindent where $P(i)$ indicates the Pauli $P$ acting on qubit $i$. The above procedure can be extended to systems with more spins. For
6 spins, the 64 possible spin configurations can be reduced to a
composite basis of eight states. The quantum computation can be done
using three qubits. The circuit for the 6-spin variational wave function
depending on seven parameters is shown in Fig. \ref{sixspin}. This circuit has been optimized for a quantum architecture with only nearest-neighbour qubit connectivity. 

\section{Simulation results}\label{sec:simres}

Computations were carried out for 4- and 6-spin systems using quantum
simulators. Code was written using the PennyLane differentiable programming
framework. For production work, the PennyLane \texttt{default.qubit} and Qiskit
Aer simulators were used (through the PennyLane-Qiskit plugin)\footnote{To check cross-platform capability, code was a tested on Cirq and
Rigetti Forest simulators also.}. The codes and data files used for this work are available on GitHub \cite{ourgithub}.

The first step in the computation was the determination of optimal
variational wave function parameters for a number of different values
of the Hamiltonian parameter $r$ spanning a range from 0.5 to 1.4 (this region was chosen as it is centered around the point of a phase transition at $r=1$).
This was done using a standard variational quantum eigensolver (VQE). 
VQE was performed using fully analytical simulation. All variational parameters were
initialized to 0. A gradient descent optimizer 
with step size 0.1 was used, with the maximum number of iterations set to 1000, however optimization would terminate early if the energy computed at a 
particular iteration was within $10^{-8}$ of the true value obtained by exact diagonalization.

\begin{figure}[t]
\centering
\includegraphics[scale=0.45,trim={4cm 10cm 1cm 10cm},clip=true,angle=0]{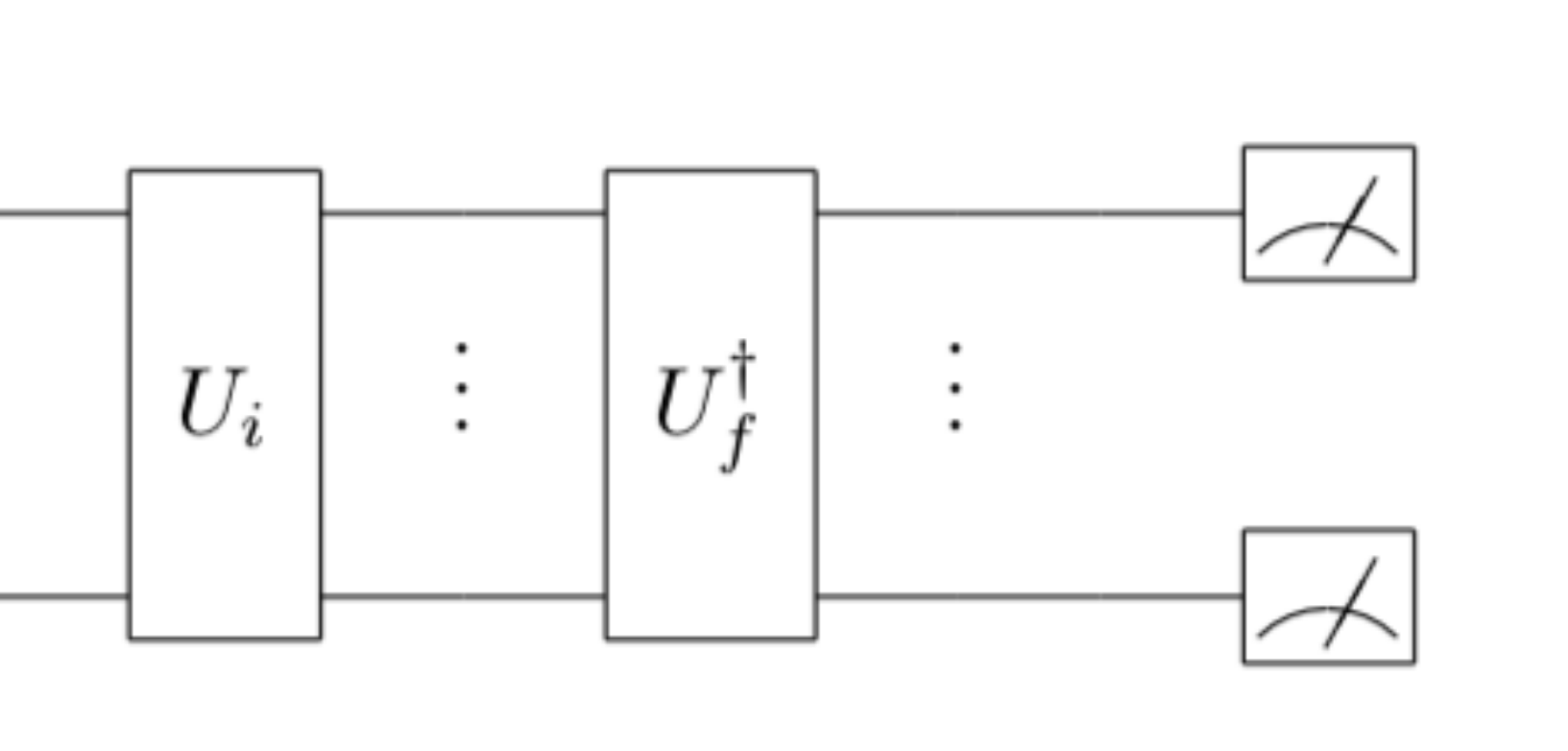}
\caption{Circuit for calculating the squared overlap of the wave functions instantiated by the
unitary operators $U_i$ and $U_f$. } 
\label{uudag}
\end{figure}

Both the fidelity susceptibility, Eq. (\ref{eq:d2over}), and 
the second derivative of the energy, Eq. (\ref{eq:d2E}), are of interest 
as indicators of a phase transition
(see Ref.\cite{PhysRevA.77.032111}). This requires computation of the three quantum gradients specified in Fig. \ref{fig:flowchart}.
PennyLane computes these automatically by applying parameter-shift rules to circuits
for the Hamiltonian expectation value and squared wave function overlap. The circuit
employed for the square of the wave function overlap is shown in generic
form in Fig. \ref{uudag}. Starting in a state of all 0's, the probability
of measuring all 0's gives the squared overlap. If one could not make
use of Eq. (\ref{eq:d2ovsqr}) and the overlap, instead of its square, 
was really required, a Hadamard test could be used. This is discussed in 
Appendix A.

\begin{figure}[t]
\centering
\includegraphics[width=\columnwidth]{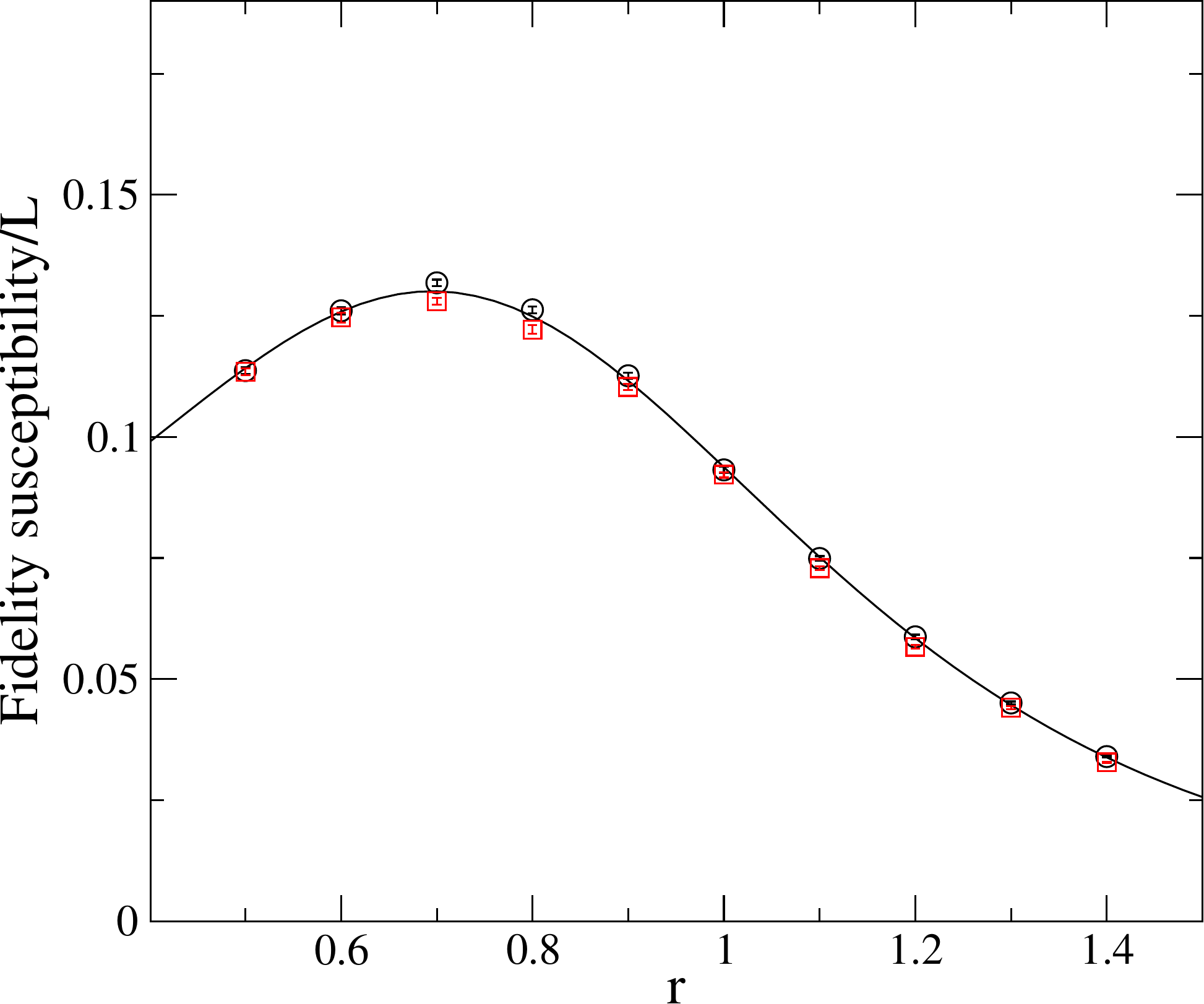}
\caption{Fidelity susceptibility per site of the transverse field Ising model as a function 
of the Hamiltonian parameter $r$ for a lattice of four sites.  The circles are for an ideal (noiseless) quantum computation. The squares are results of a noisy simulation. The solid
lines are exact results.} 
\label{exactFS4}
\end{figure}

\begin{figure}[H]
\centering
\includegraphics[width=\columnwidth]{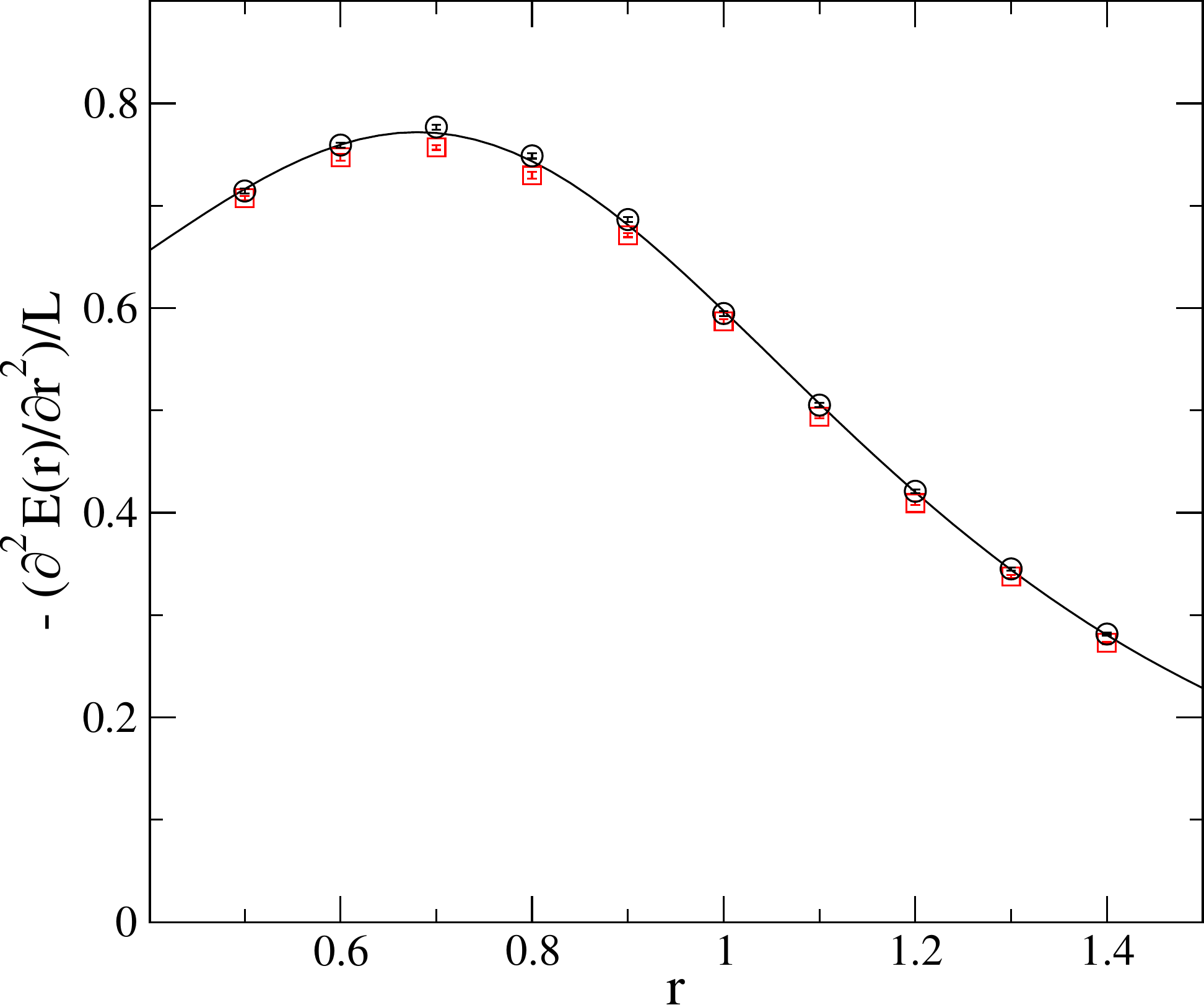}
\caption{Second derivative of the energy per site of the transverse field Ising model as a function 
of the Hamiltonian parameter $r$ for a lattice of four sites.The circles are for an ideal (noiseless) quantum computation. The squares are results of a noisy simulation. The solid
lines are exact results.} 
\label{exactSED4}
\end{figure}

\begin{figure}[t]
\centering
\includegraphics[width=\columnwidth]{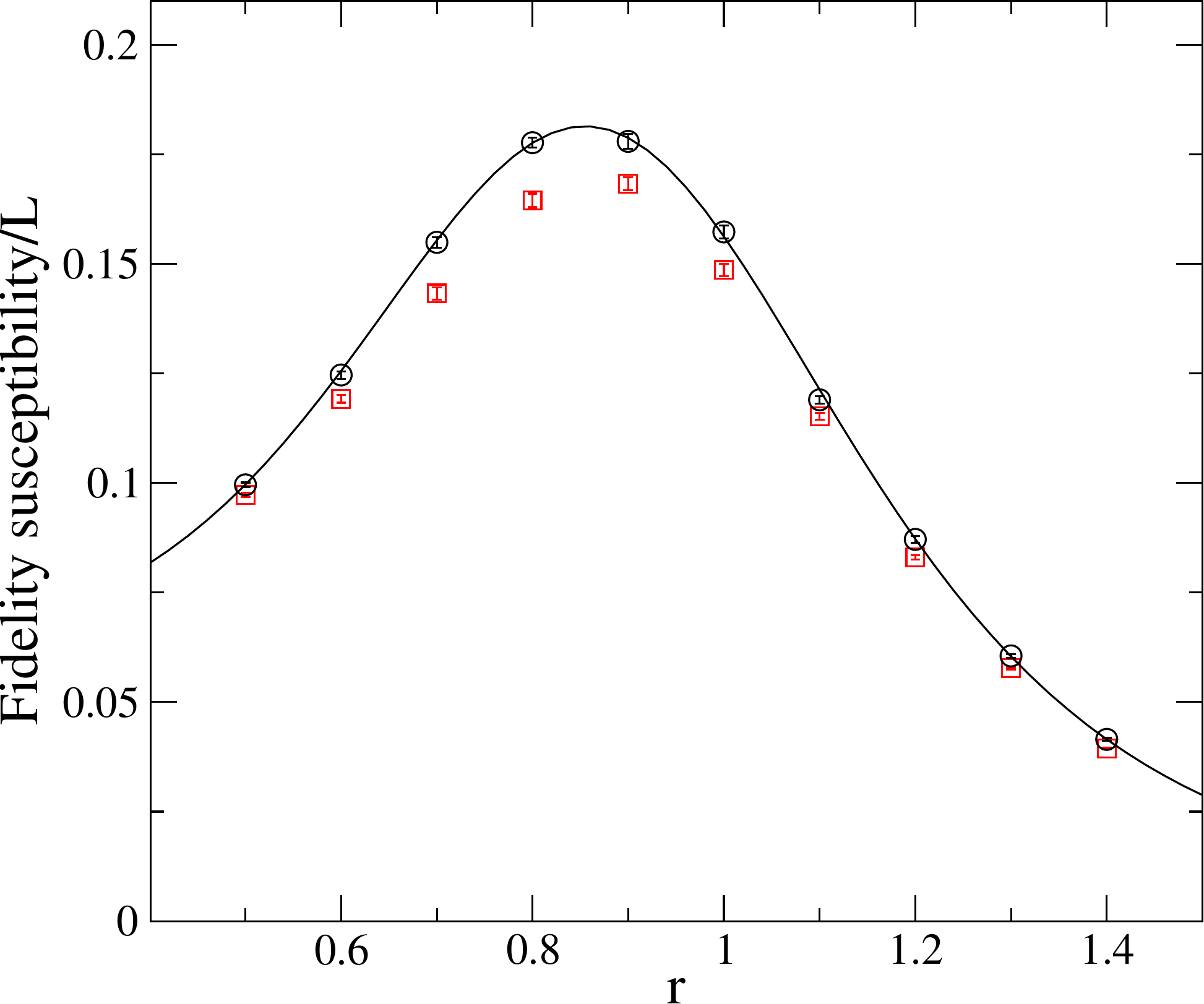}
\caption{Fidelity susceptibility per site of the transverse field Ising model as a function 
of the Hamiltonian parameter $r$ for a lattice of six sites. The circles are for an ideal (noiseless) quantum computation. The squares are results of a noisy simulation. The solid
lines are exact results.} 
\label{exactFS6}
\end{figure}

\begin{figure}[hbt]
\centering
\includegraphics[width=\columnwidth]{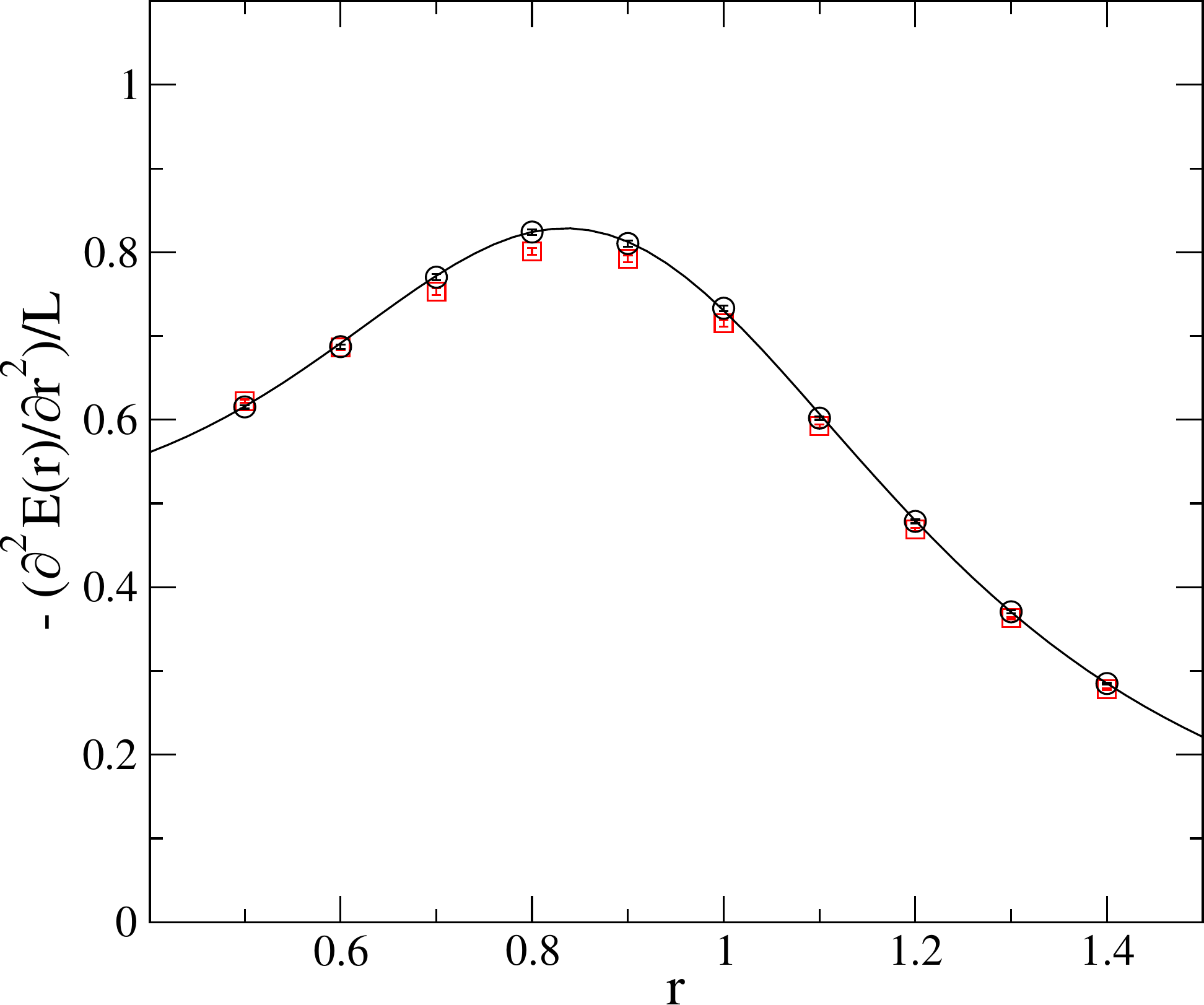}
\caption{Second derivative of energy per site of the transverse field Ising model as a function 
of the Hamiltonian parameter $r$ for a lattice of six sites.  The circles are for an ideal (noiseless) quantum computation. The squares are results of a noisy simulation. The solid
lines are exact results.} 
\label{exactSED6}
\end{figure}

To verify that automatic differentiation was working properly the fidelity
susceptibility (\ref{eq:d2over}) and the second derivative of the energy  
(\ref{eq:d2E}) were computed as a function of the Hamiltonian 
parameter $r$ using the PennyLane \texttt{default.qubit} device 
which simulates an ideal (i.e., noise-free) quantum computer. The circles in Figs. \ref{exactFS4} and \ref{exactSED4}
show the results, averaged 
over twenty trials using 8192 shots for each measurement per trial, for spin chains of length 4. 
The solid lines are the results obtained by exact solution of the model. The corresponding results for a lattice of six sites are shown in Figs. \ref{exactFS6} and \ref{exactSED6}.

On a real quantum device, gate errors and readout errors will degrade the
accuracy of the computation. Readout errors can be mitigated in a fairly
straightforward way so in this study these effects were not
included. The focus here is in quantifying gate error effects and the 
mitigation of gate errors is discussed in the next section. The
Qiskit Aer simulator is used since it provides access to
noise models for actual (IBM Q) hardware devices.


\
The four spin computation is done on two qubits using publicly available
calibration data, so we can choose a device which has the minimum 
CNOT gate error. At the time the computation was done, this was the \texttt{ibmq\_manila}
device. The results of noisy simulations averaged over twenty trials are shown with square symbols in Figs. \ref{exactFS4} and \ref{exactSED4}. 
The results were calculated using the qubit pair with the largest CNOT error. For this calculation, where the wave function ansatz contains only a
single two-qubit gate, even this least favourable choice shows little sensitivity to gate errors.

The six spin model, even simplified to run on three qubits, is more challenging. In order to entangle all qubits more CNOT gates are required. 
We choose to use CNOT gates only between neighbouring qubits 
(Fig. \ref{sixspin})
rather than relying on transpiler-generated swap operations. This is to facilitate the error mitigation study discussed in the next section.
The simulation results using
the \texttt{ibmq\_manila} noise model and coupling map, which specifies connectivity, are plotted with square symbols in Figs. \ref{exactFS6} and \ref{exactSED6}. Note that no transpiler 
optimization was applied in doing these calculations.

\section{Error mitigation and automatic differentiation}
\label{sec:error_mitigation}

\subsection{Richardson extrapolation}
\label{subsec:richardson}

Zero-noise extrapolation is a common means of performing error mitigation 
on the results of quantum computations run on hardware. It involves
scaling up the amount of noise in a computation through the systematic addition of noisy gates, followed by extrapolation back to a noise-less value. 
Richardson extrapolation (RE) is a commonly-applied technique \cite{temme2017error, krebsbach2022optimization}; it is based on
polynomial interpolation using \emph{Lagrange basis polynomials}. Suppose we
measure a set of expectation values at $n + 1$ different points (e.g.,
different noise scale factors), labelled $(x_j, E_j)$, for $j=0,
\ldots, n$. The Lagrange polynomial for this set of points is the lowest-degree
polynomial that passes through all of them exactly. It can be constructed as a linear combination of the expectation values and Lagrange basis polynomials,
\begin{equation}
P_n (x) = \sum_{j=0}^{n} E_j \ell_j(x),
\end{equation}
\noindent where 
\begin{equation}
    \ell_j(x) = \prod_{\substack{0 \leq m \leq n, \\ m \neq j}} \frac{x - x_m}{x_j - x_m}.
\end{equation}

\noindent To perform error mitigation with RE, first construct the Lagrange polynomial,  then evaluate it at $x = 0$ to obtain an estimate for the mitigated value. A somewhat simplified expression of the polynomial can be used,
\begin{eqnarray}
    P_n (0) = \sum_{j=0}^{n} E_j \ell_j(0) 
    &=&  \sum_{j=0}^{n} E_j  \prod_{\substack{0 \leq m \leq n, \\ m \neq j}} \frac{ - x_m}{x_j - x_m} \nonumber \\
    &=& \sum_{j=0}^{n} E_j \gamma_j, \label{eq:richardson_extrap}
\end{eqnarray}
as per the notation in Eqs. (3) and (4) of
\cite{krebsbach2022optimization}. From here on we will denote this value by
$\mathcal{E}$ (mitigated values will in general be represented using a calligraphic font).

In this work, we are interested in performing error mitigation not only on expectation values, but also on gradients of them. Suppose that $\theta$ is the parameter with respect to which we would like to compute a derivative.  It is straightforward to show that for $\mathcal{E} = P_n(0)$,
\begin{equation}
    \frac{\partial \mathcal{E}}{\partial \theta} = \frac{\partial}{\partial \theta} \left( \sum_{j=0}^{n} E_j \gamma_j \right) = \sum_{j=0}^{n} \frac{\partial E_j}{\partial \theta} \gamma_j,
\end{equation}
\noindent and so the gradient of an error-mitigated value is simply the error-mitigated computation of a gradient. Similarly, second derivatives are required, and work the same way:
\begin{equation}
    \frac{\partial^2 \mathcal{E}}{\partial \theta^2} = \sum_{j=0}^{n} \frac{\partial^2 E_j}{\partial \theta^2} \gamma_j. \\
\end{equation}

\noindent Thus, computing gradients and performing error mitigation within the larger context of automatic differentiation will in theory commute. 

When we apply RE in practice, each $E_j$ in Eq. (\ref{eq:richardson_extrap})
has been computed independently based on $N_j$ shots. What is obtained, then, is an estimate of $\mathcal{E}$, $\hat{\mathcal{E}}$, computed as a linear combination of estimates $\hat{E}_j$,
\begin{equation}
    \hat{\mathcal{E}} = \sum_{j=0}^{n} \hat{E}_j \gamma_j.
\end{equation}

\noindent The variance of the mitigated value, $\sigma^2$, can be computed in terms of the
variances of the individual estimates. Usually the variance of a sum of terms
involves covariance terms. However, it is reasonable to assume measurements of the expectation values are independent so all
covariances are 0. This yields
\begin{equation}
    \sigma^2 = \sum_{j=0}^{n} \sigma^2_j \gamma_j^2. \label{eq:var_initial}
  \end{equation}

In the broader context of a quantum algorithm, such as the one considered here, 
one is computing the expectation value of not only a
single observable, but of a full Hamiltonian. Let
\begin{equation}
 H = \sum_{i} c_i P_i, 
\end{equation}

\noindent be the Hamiltonian of interest, where $P_i$ are $n$-qubit
Pauli operators. The error-mitigated estimate of the energy, $\hat{\mathcal{H}}$,
is given by
\begin{equation}
 \hat{\mathcal{H}} = \sum_{i} c_i \hat{\mathcal{P}}_i.
\end{equation}
\noindent If each estimate $\hat{\mathcal{P}}_i$ has variance $\sigma^2_{P_i}$
according to Eq. (\ref{eq:var_initial}), then 
the total variance is
\begin{equation}
 \sigma_H^2 = \sum_{i} c^2_i \sigma^2_{P_i} = \sum_i  c_i^2 \sum_{j=0}^n (\sigma^2_{P_i})_{j}\gamma_j^2. 
\end{equation}

\noindent The variance of error-mitigated estimates 
is higher than those of the unmitigated values, a fact which has been remarked upon in a number of previous works \cite{krebsbach2022optimization, wang2021error, qermit}. If ``too much" error mitigation is applied (e.g., if $n$ is too large), the results may be worse than an unmitigated value unless compensated for by increasing the number of shots to reduce the error in the estimated quantities; this is elaborated on in Appendix B. The quality of mitigation can be quantified using absolute mitigation error  \cite{qermit}, which is reported for these simulations in Appendix C. 

The increase in variance is further compounded by the fact that mitigated values of first and second derivatives are being computed, each of which comprises a sum of multiple shifted terms. For example, for the gradient of a single expectation value, 
\begin{eqnarray}
    \frac{\partial \hat{\mathcal{E}}}{\partial \theta} &=& \sum_{j=0}^{n} \frac{\partial \hat{E}_j}{\partial \theta} \gamma_j \nonumber \\ &=& \sum_{j=0}^{n} \frac{1}{2} \left(\hat{E}_j (\theta+\pi/2) - \hat{E}_j(\theta - \pi/2) \right) \gamma_j.
\end{eqnarray}

\noindent The variance of the estimate is
\begin{equation}
    (\sigma^2)^\prime = \sum_{j=0}^{n} \frac{1}{4} (\sigma^2_{j+} + \sigma^2_{j-}) \gamma_j^2 \approx \sum_{j=0}^{n} \frac{1}{2} (\sigma^2_{j+}) \gamma_j^2
\end{equation}

\noindent where $\sigma^2_{j\pm}$ represents the variance of the $j^{th}$ term at a shifted value of $\pm \pi/2$. It is assumed that, due to symmetry, these variances will be equal (but note that they are not necessarily equal to the variance at the unshifted value) \cite{mari2021estimating}. Thus, for the derivative of the expectation value of a full Hamiltonian, a mitigated estimate is obtained with variance 
\begin{equation}
 (\sigma_H^2)^\prime = \sum_{i} c^2_i (\sigma^2_{P_i})^\prime = \sum_i  c_i^2 \sum_{j=0}^{n} \frac{1}{2} (\sigma^2_{P_i})_{j+} \gamma_j^2.
\end{equation}

\noindent A similar expression can be derived for Hessians (this is relegated to Appendix B). In both cases there is a dependence on the parameter-shift rule used, and in particular on the number of terms it contains. The increased variance has consequences in our problem due to where they are used: in the solution of the response equations, Eq. (\ref{eq:first_order_response}), whose results are propagated into the computations of second energy derivative and fidelity susceptibility. Even small deviation can result in the system of equations being poorly conditioned, leading to suboptimal solutions.

\subsection{Software implementation}

 While we note that the open-source error-mitigation library Mitiq \cite{mitiq} has a great variety of error mitigation functionality, mitigation was implemented using the differentiable quantum transforms framework in PennyLane \cite{dimatteo2022transforms} in order to preserve differentiability. Transforms are composable
 ``metaprograms" that modify the behaviour of quantum functions and circuits. 
In particular, we leverage the concept of a \emph{batch transform}: such transforms take quantum functions as input, and return a set of transformed functions, and a processing function that acts on the results of those functions.  This is shown below in pseudocode:

\begin{lstlisting}[language=python]
transformed_fns, processing_fn = batch_transform(fn)

results = execute(transformed_fns)

processing_fn(results)
\end{lstlisting}

Expansion of a Hamiltonian into Pauli terms, computation of gradients and Hessians, and zero-noise extrapolation can all be expressed as batch transforms.  For example, the gradient computation as described in Sect.~\ref{subsec:quantum_gradients} is a batch transform: a quantum circuit is passed as input, and the transform returns two quantum circuits as output (one for each shifted parameter), and a function that combines the executed results (Eq.~(\ref{eq:param_shift})). Furthermore, if the extrapolation procedure is implemented in an autodifferentiable way, the error-mitigated results can be input to subsequent computations without compromising differentiability. Pseudocode for the core function to compute mitigated gradients is shown below to highlight key elements of batch transform functionality such as composition and use of the reconstruction functions. The full implementation is included in our GitHub repository \cite{ourgithub}.

\begin{lstlisting}[language=python]  
import pennylane as qml

# A quantum device 
dev = qml.device(...)

# A quantum circuit that computes the 
# expectation of a Hamiltonian that we 
# would like to obtain an error-mitigated 
# gradient for
circuit = ...

# Some configuration options for error
# mitigation; passed to ZNE routine
mitigation_config = {...}

# Split into constituent terms so mitigation 
# is applied to each one separately
ham_circuits, ham_fn = qml.transforms.hamiltonian_expand(circuit, group=False)

# Map gradient computation batch transform 
# over each circuit from the previous step
grad_circuits, grad_fn = qml.transforms.map_batch_transform(
    qml.gradients.param_shift, ham_circuits
)

# Map each gradient circuit into a set of 
# circuits that computes the error-mitigated
# value. This set of circuits is then  
# executed by the quantum device. 
zne_circuits, zne_fn = qml.transforms.map_batch_transform(
    partial(zne, mitigation_config), grad_circuits,
)

execution_results = qml.execute(
    zne_circuits, dev, gradient_fn=None
)

# The processing functions keep track 
# of what should happen to each
# value during the transform, and are 
# used to recombine the results
error_mitigated_expval_gradient = ham_fn(grad_fn(zne_fn(execution_results))
)

\end{lstlisting}

The \texttt{mitigation\_config} parameter in the pseudocode  above 
provides a great deal of choice in how error mitigation is performed. For instance, how many 
noise scale factors are used, what their values are, and how noise gets added 
based on them. This work considers two ways of adding noise: unitary folding \cite{giurgicatiron2020digital} and CNOT folding \cite{he2020zeronoise}, each of which are implemented as quantum function transforms \cite{dimatteo2022transforms}. In unitary folding, given an input circuit $U(\theta)$ and an integer noise scale factor $\lambda_j$, the circuit is transformed into $(U(\theta) U^\dagger(\theta))^{(\lambda_j-1)/2} U(\theta) $, i.e., the circuit is run forwards and backwards $(\lambda_j-1)/2$ times. In CNOT folding $(\lambda_j-1)/2$ pairs of CNOTs are added after each existing CNOT.

To verify the mitigation pipeline, it was first applied to computation of ground state energy, the second energy derivative, and the fidelity susceptibility by running the circuit at the optimal variational parameters with no noise, as well as with shot noise. Following this, the simulated device noise model of \texttt{ibmq\_manila} was applied, with the circuits adjusted for the coupling map prior to any transforms being applied. Simulations were run on a desktop machine, but due to the extensive run time of noisy simulations, execution was parallelized over the different values of $r$. 

We experimented with sequences of noise scale factors with varying length. Given $\lambda_j = 2j + 1$, simulations were run for sequences of scale factors $[\lambda_0, \ldots, \lambda_n]$ for $n = 0$ (i.e., noisy simulation only) up to $n=4$ (i.e., a max $\lambda_j$ of $9$) for $L=4$, and up to $n=3$ for $L=6$. No notable difference was observed between CNOT folding and unitary folding. In both cases, as the number of folds increases, so does the variance in results; however, due to the nature of the quantities being computed, this variance was strongly amplified, most notably for the cases of max $n=3$ and $n=4$ (up to 3 and 4 folds, respectively).

\begin{figure*}[ht]
    \centering
    \includegraphics[width=\textwidth]{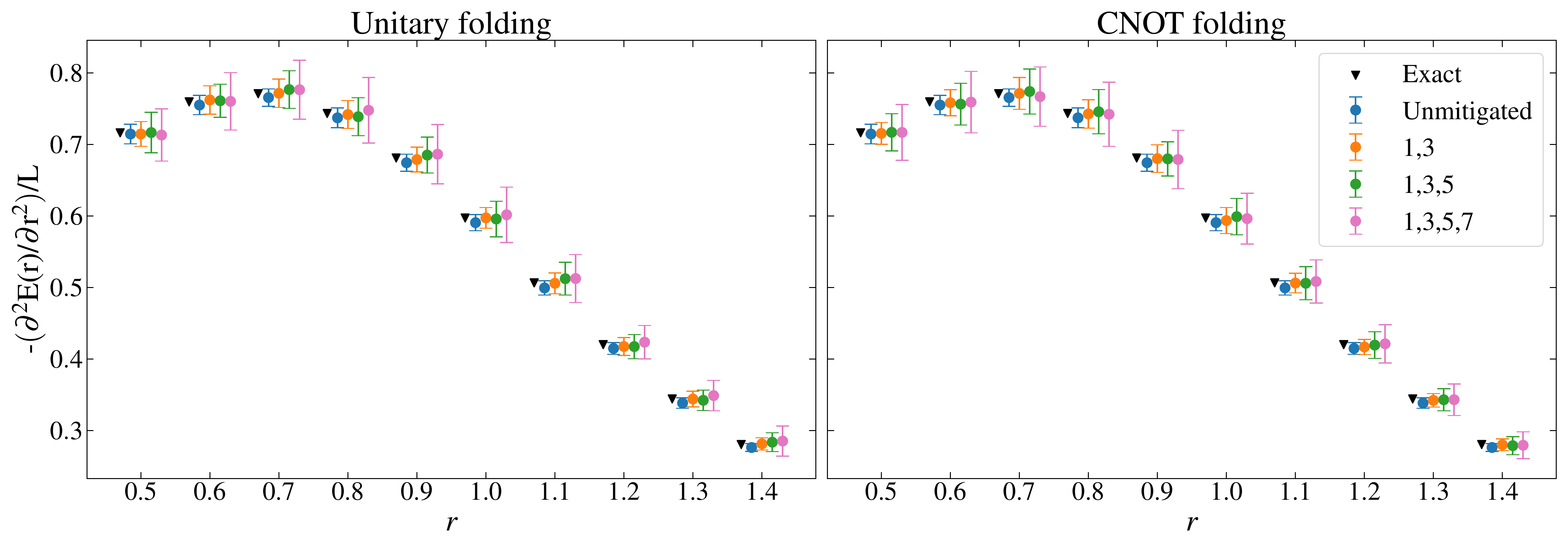}
    \caption{Second energy derivative per site for the 4-spin system. Results are plotted along the horizontal axis for $r \in [0.5, 0.6, \ldots 1.4]$ but offset for readability.  The numbers in the legend indicate which scale factors were used in the mitigation process (e.g., ``1,3,5" means mitigation was performed using 3 points for the extrapolation: the unmodified circuit, adding one fold, and adding two folds). For each mitigation method, 100 independent trials were run; the error bars correspond to the standard deviation of these results. As expected, a greater number of points in the extrapolation leads to higher variance in the potential result.}
    \label{fig:l4_d2e}
\end{figure*}

\begin{figure*}[ht]
    \centering
    \includegraphics[width=\textwidth]{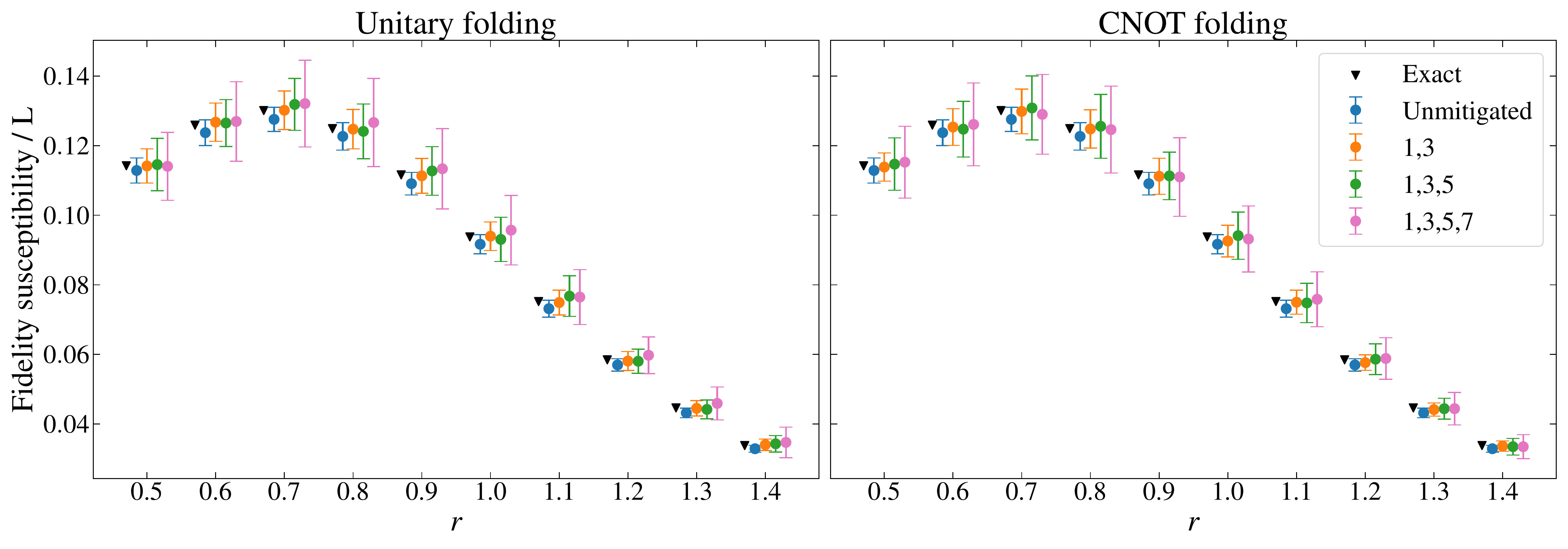}
    \caption{Fidelity susceptibility per site for the 4-spin system. For each mitigation method, 100 independent trials were run; the error bars correspond to the standard deviation of these results. Greater variation is observed, compared to the second energy derivative of Fig. \ref{fig:l4_d2e}, due to a double dependence on the derivatives of variational parameters with respect to the Hamiltonian parameter $r$.}
    \label{fig:l4_fs}
\end{figure*}

\begin{figure*}[ht]
    \centering
    \includegraphics[width=\textwidth]{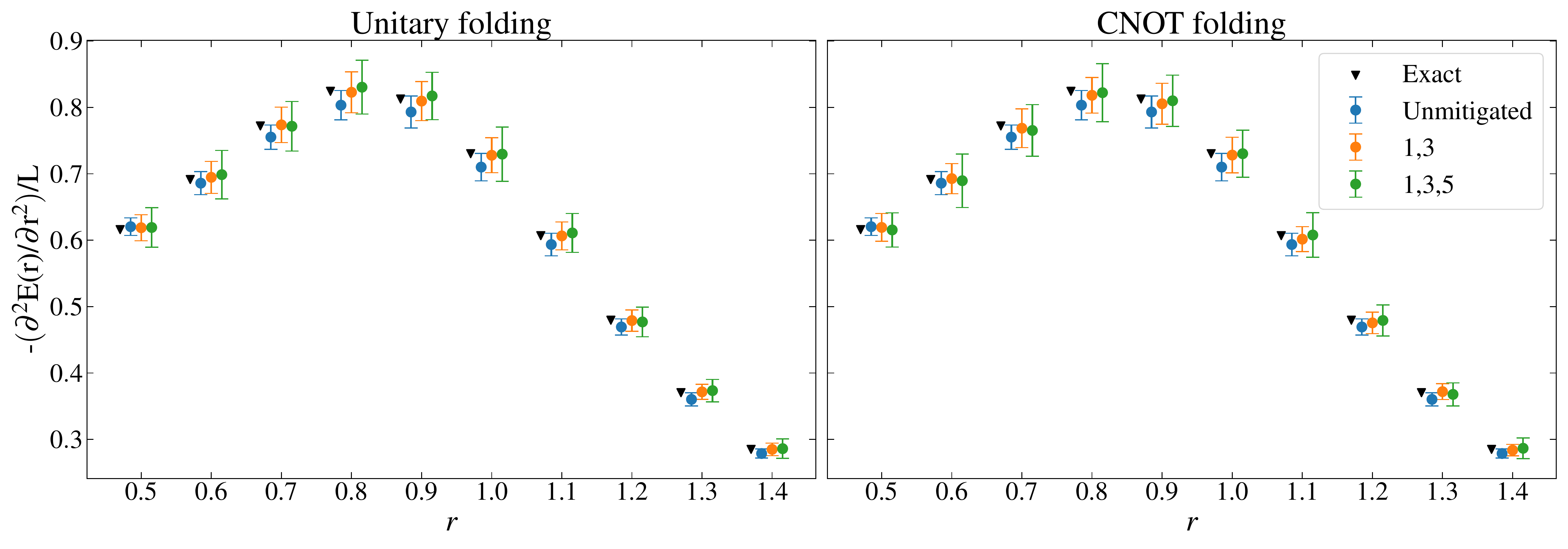}
    \caption{Second energy derivative per site for the 6-spin system. Error bars correspond to the standard deviation of 100 independent trials. Compared to results in the 4-spin case, the variance is higher for both types of folding.}
    \label{fig:l6_d2e}
\end{figure*}

\begin{figure*}[ht]
    \centering
    \includegraphics[width=\textwidth]{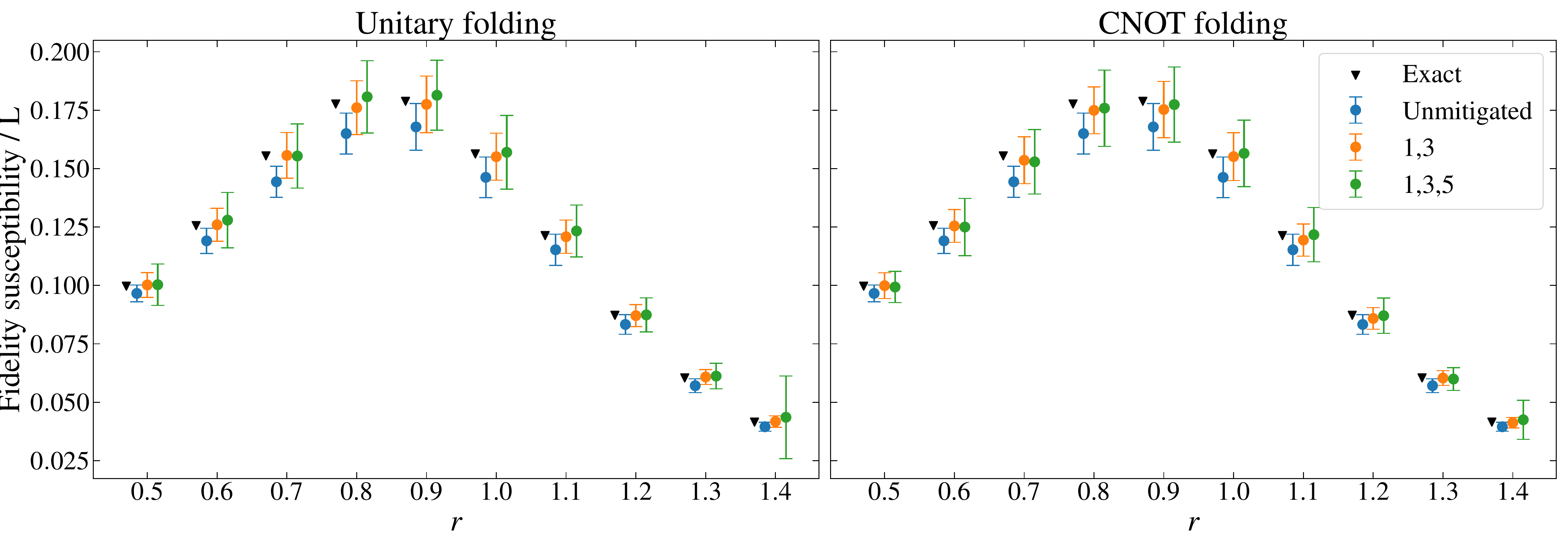}
    \caption{Fidelity susceptibility per site for the 6-spin system. Error bars correspond to the standard deviation of 100 independent trials. Similar to the 4-spin case, the variance in results is higher than that of the second energy derivative. In the case of unitary folding at $r=1.4$, one trial resulted in a markedly large condition number for the matrix inversion involved in solution of the response equations, leading to significantly less accurate results. We found such behaviour to be common for the 6-spin case when pushing beyond the use of three scale factors for both types of folding.}
    \label{fig:l6_fs}
\end{figure*}

Results are shown here for up to the $n=3$ case for $L=4$, and up to $n=2$ for $L=6$; the full set of numerical results is available for perusal on GitHub \cite{ourgithub}. 
The $L=4$ ($L=6$) results are shown in Fig. \ref{fig:l4_d2e} and \ref{fig:l4_fs} (Figs. \ref{fig:l6_d2e} and \ref{fig:l6_fs}). To obtain the error-mitigated estimate 8192 shots were used for \emph{each circuit} that was run. The entire computation was run 100 times for each quantity of interest; the error bars on the plots show the mean and standard deviation of the results. We observe the clear increase in variance which occurs from performing error mitigation as the number of points used in the extrapolation is increased. Note that, when results are averaged over all error-mitigated attempts, we do in many cases obtain a mitigated value that improves over the unmitigated one; however this requires a substantial amount of quantum resources. This is further discussed in Appendix C, where plots are provided for the absolute mitigation error over the distribution of results.

In some trials, a significant variation was observed, most notably for simulations with a larger number of folds and scale factors. It was found that the reason for this was due to the dependence of both the second energy derivative and the fidelity susceptibility on the quantities $\partial \theta_i / \partial r$. Error mitigation was applied to the computation of each element of the Hessian matrix and gradient of the expectation value of $H_1$ required as per the response equations (Eq. (\ref{eq:first_order_response})). Their solution involves a linear inversion, but when the Hessian is noisy the resulting matrix may be poorly conditioned, leading to very large values of the $\partial \theta_i / \partial r$. These are then used directly in computation of the second energy derivative, and \emph{twice} in the computation of fidelity susceptibility (so its variance is even higher). Obtaining accurate values of these parameters is thus critical.

\section{Summary}

In this work an end-to-end implementation of quantum automatic differentiation for a problem in condensed matter physics was presented. The use of automatic differentiation enables the computation of quantities of interest (second energy derivative and fidelity susceptibility) using effectively a single quantum circuit, without knowledge of the full spectrum of the system. It is hoped that this demonstration will motivate further exploration into how these tools can be applied to other physical problems.

The methods were successfully applied here to small quantum systems with correct results, up to deviations due to statistical and simulated hardware noise. However, the computation of error-mitigated gradients with respect to Hamiltonian parameters was found to be particularly sensitive to such noise. A more general and systematic study should be performed to analyze the theory underlying error-mitigation of energy derivatives, as it has implications for many other problems in areas such as quantum chemistry. This could include testing different methods for computing the matrix inverse when solving the response equations (e.g., through a pseudoinverse / singular value decomposition), or alternative strategies for evaluating the derivatives themselves \cite{obrien2019calculating, Azad2022}. Different (non-ZNE) error mitigation techniques should also be analyzed, and it would be of interest to determine the extent to which the mitigation strategy itself could be optimized, using, e.g., methods developed in \cite{krebsbach2022optimization} or \cite{vaqem}. Finally, as noted in \cite{vaqem}, results of error mitigation applied to noisy simulations may not be representative of those obtained using an actual quantum processor, and so, if feasible, future work should include execution on real devices. 

\section{Acknowledgements}

ODM thanks Josh Izaac and Tom Bromley for helpful discussions, and acknowledges funding provided by NSERC grant no. RGPIN-2022-04609. TRIUMF receives federal funding via a contribution agreement with the National Research Council of Canada. We acknowledge the use of the IBM Qiskit framework for
this work. The views expressed are those of the authors,
and do not reflect the official policy or position of IBM
or the IBM Quantum team.

\section*{Appendix A: Overlap algorithms}

\begin{figure}[ht]
\centering
\includegraphics[width=\columnwidth]{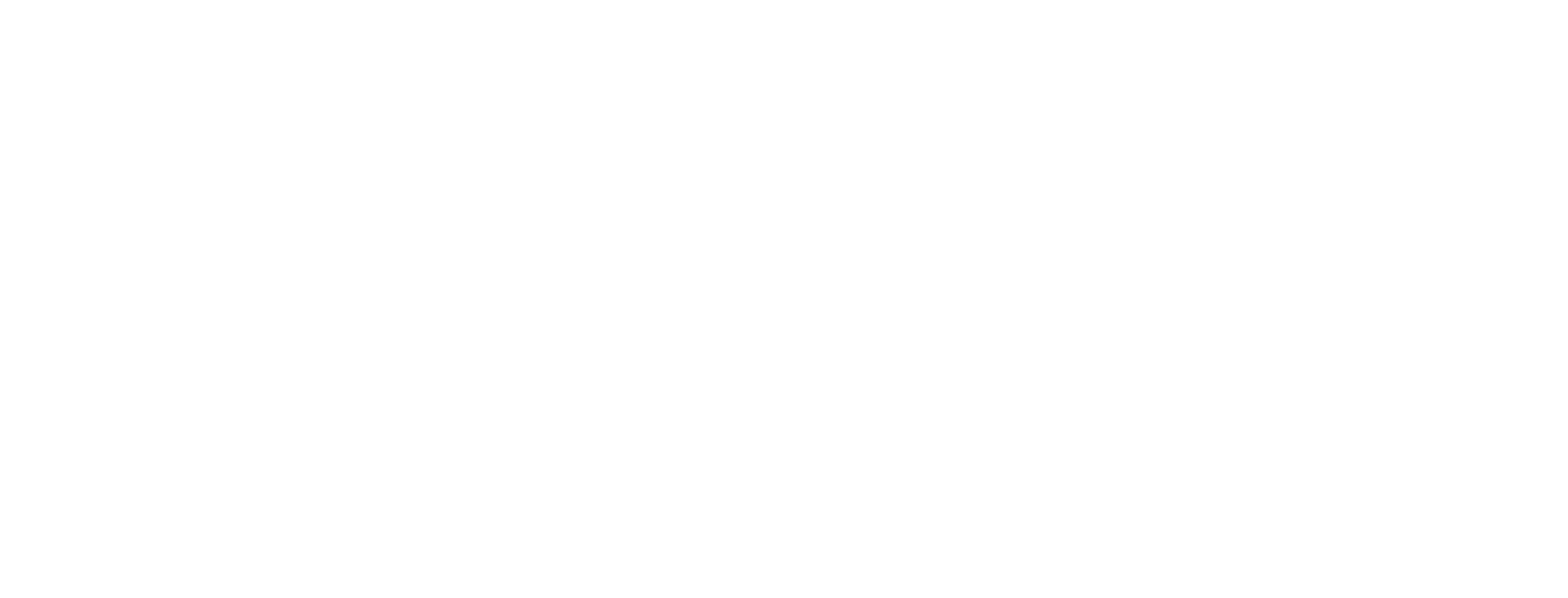}
\caption{Circuit for calculating the overlap of the wave functions instantiated by the
unitary operators $U_i$ and $U_f$ using the Hadamard test. The gate R is R${_y}(-\pi/2)$ or 
R${_x}(\pi/2)$ for the real and imaginary parts respectively.} 
\label{hadam}
\end{figure}

\begin{figure}[ht]
\centering
\includegraphics[width=\columnwidth]{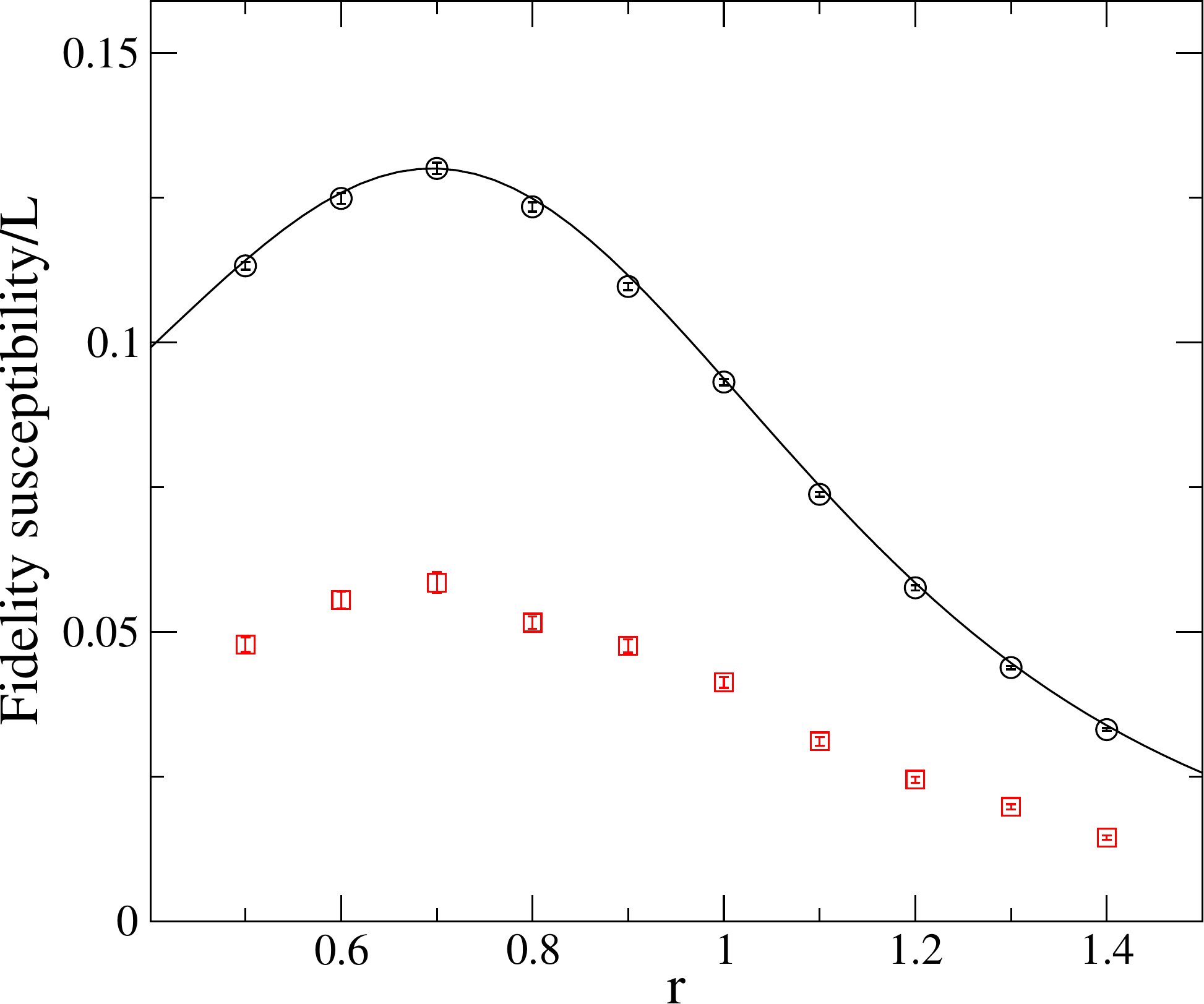}
\caption{Fidelity susceptibility of the transverse field Ising model as a function of the 
Hamiltonian parameter $r$ for four spins comparing the Hadamard test  (squares)with the overlap 
circuit Fig. \ref{uudag} (circles). The solid line is the exact result.} 
\label{hadFS}
\end{figure}

The Hadamard test is a standard way to calculate the overlap between two states.
The circuit for this test is shown in generic form in Fig. \ref{hadam}. The advantage of
the Hadamard test is that it can provide separately the real and imaginary parts
of the overlap. The disadvantages, compared to the circuit in Fig. \ref{uudag}
are that it uses an additional qubit and, more importantly, it requires
controlled versions of the unitary operators that create the states. This leads
to a large increase in the number of multi-qubit gates resulting in greater
sensitivity to hardware noise. This is illustrated in Fig. \ref{hadFS} where the fidelity
susceptibility as a function of the Hamiltonian parameter $r$ for 4 spins calculated
using the Hadamard test, and is compared to that calculated with overlap from 
Fig. \ref{uudag}. The simulation is done using the \texttt{ibmq\_manila} gate noise model. 

\begin{figure}[ht]
\centering
\includegraphics[width=\columnwidth]{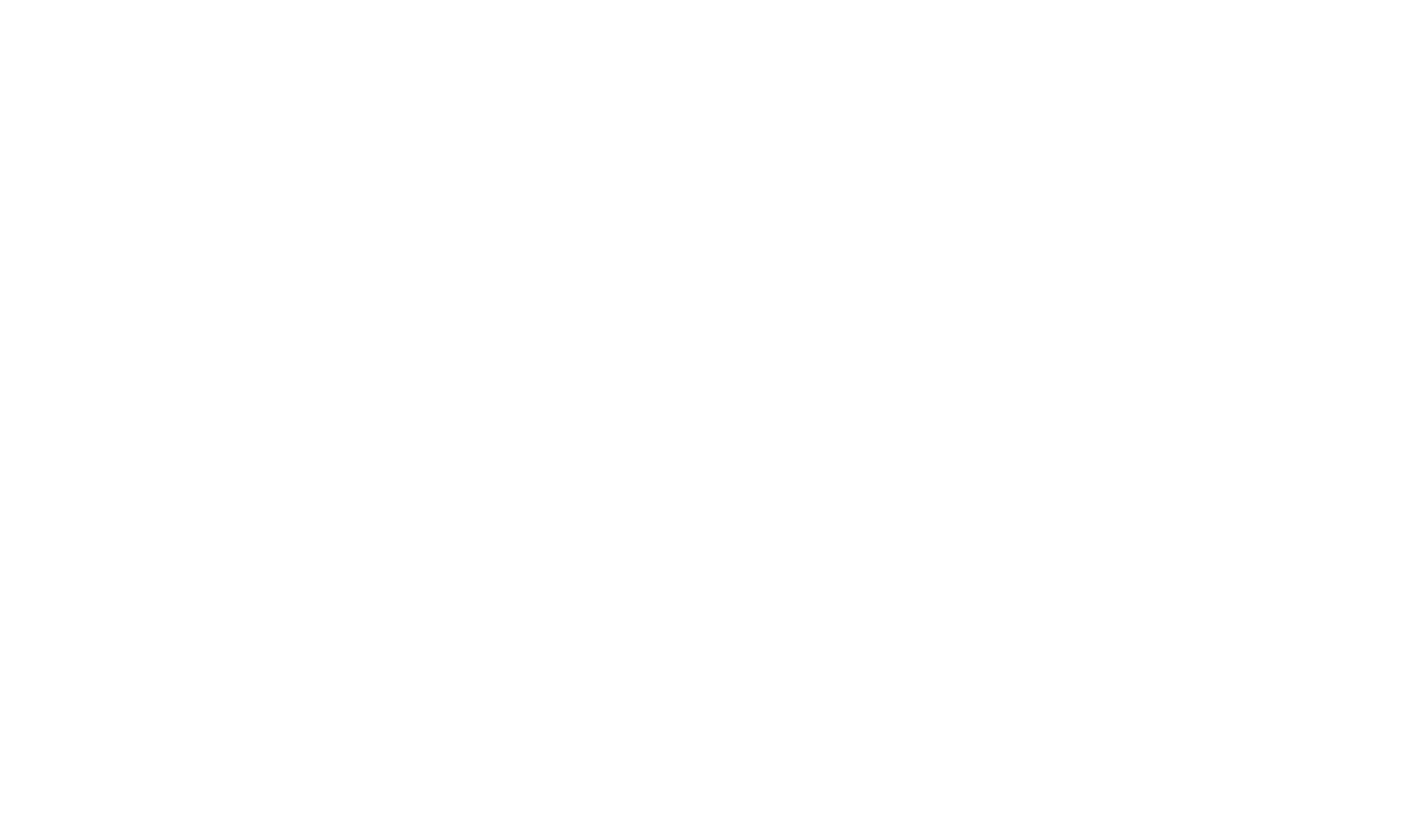}
\caption{Circuit for calculating the squared overlap of the wave functions instantiated by the
unitary operators $U_i$ and $U_f$ using the swap test. } 
\label{swap}
\end{figure}

\begin{figure}[ht]
\centering
\includegraphics[width=\columnwidth]{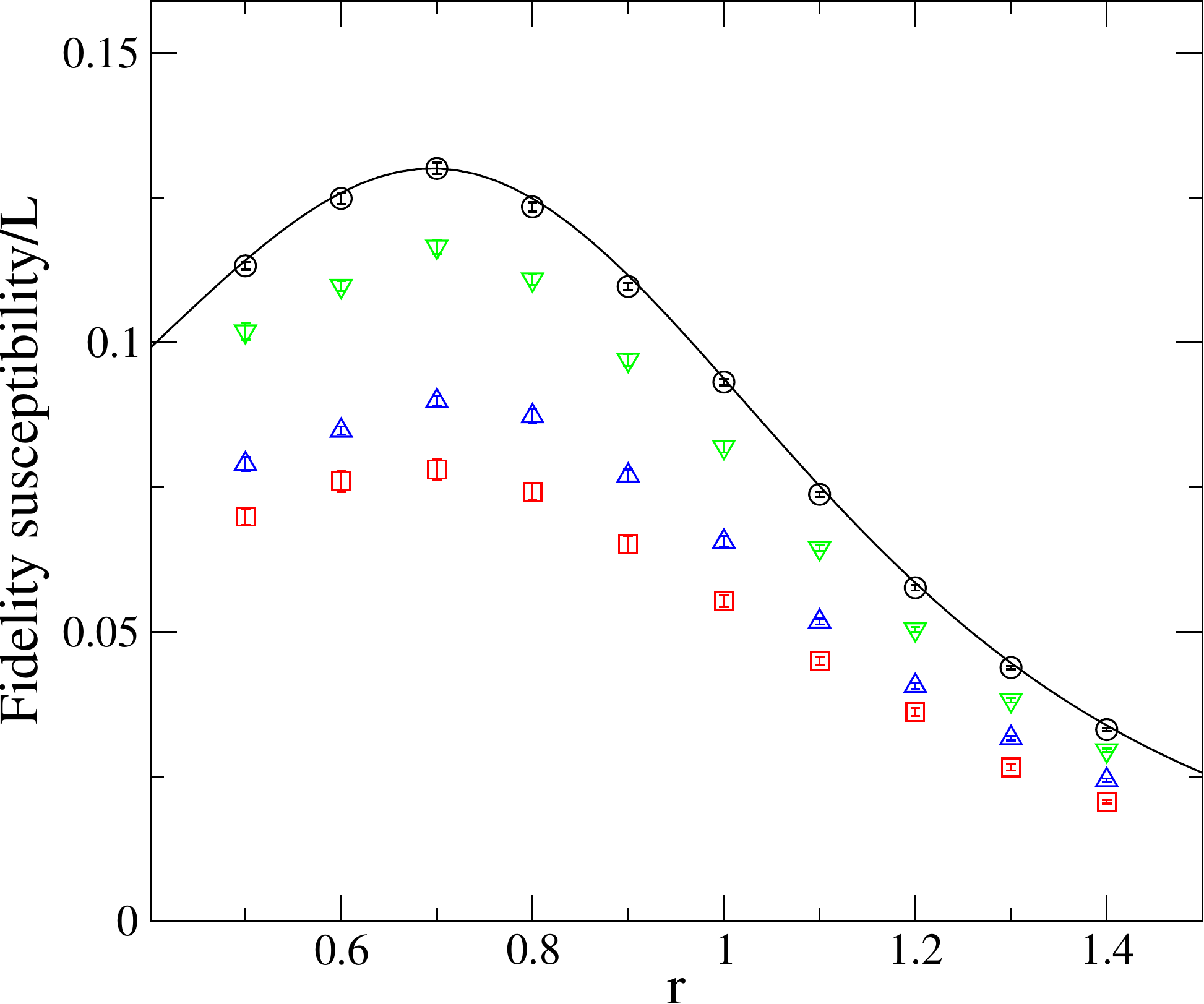}
\caption{Fidelity susceptibility of the transverse field Ising model as a function of the Hamiltonian parameter $r$ for
four spins using the swap test Fig. \ref{swap} (squares) compared to the overlap circuit 
Fig. \ref{uudag} (circles). Results of using alternate versions of the swap test discussed 
in \cite{Cincio_2018} are shown using up-triangles  (Ancilla-Based Algorithm, Ref. \cite{Cincio_2018}, Fig. 5)
and down-triangles (Bell-Based Algorithm, Ref. \cite{Cincio_2018}, Fig. 6).
The solid line is the exact result.} 
\label{swapFS}
\end{figure}

The swap test is another way to compute the squared overlap of two states. The generic circuit is
shown in Fig. \ref{swap}. The advantage of the swap test over the Hadamard test is that the unitary
operators creating the states are not controlled. However, the number of qubits required to contain 
the states is doubled. Instead of controlled unitaries, controlled swaps are needed which is a
disadvantage compared to the overlap circuit in Fig. \ref{uudag}.

In Ref. \cite{Cincio_2018} Cincio \emph{et al.} discuss ways to improve the swap test. 
Fig. \ref{swapFS} compares 
the fidelity susceptibility as a function of the Hamiltonian parameter 
$r$ for 4 spins calculated with the \texttt{ibmq\_manila} gate noise model using 
the Ancilla-Based Algorithm (ABA, Ref. \cite{Cincio_2018}, Fig. 5) and the 
Bell-Based Algorithm (BBA, Ref. \cite{Cincio_2018}, Fig. 6) with results using a 
simple unimproved swap, Fig. \ref{swap}, and the overlap circuit in Fig. \ref{uudag}.
The algorithms from Ref. \cite{Cincio_2018} provide an improvement but, as one might expect,
the circuit Fig. \ref{uudag} which does not introduce controlled gates, other than those contained 
in the state-creating unitaries, shows the least sensitivity to hardware noise. 

\section*{Appendix B: Additional details on error-mitigated gradients}
\label{appendix:sensitivity}

In Sec. \ref{subsec:richardson}, an expression was derived for how the variance of the gradient of an expectation value is inflated when error mitigation is applied. In this appendix the derivation is extended to the Hessian, and the number of shots required to counteract the increase is estimated.

We begin by expressing a general $r$-dependent Hamiltonian as 
\begin{eqnarray}
H(r) = \sum_{k=0}^{K-1} c_k P_k &=&  H_0 + r H_1\nonumber  \\
&=& \sum_{a=0}^{A-1} c_a P_a + r \sum_{b=0}^{B-1} c_b P_b.
\end{eqnarray}

\noindent Recall that to obtain the fidelity susceptibility and second energy derivative, a gradient vector and two Hessians must be estimated, as depicted in Fig. \ref{fig:flowchart}. Using parameter-shift rules, the gradients are\begin{eqnarray}
    \frac{\partial  \langle \hat{H}_1 \rangle}{\partial \theta_i} &=&\sum_{b=0}^{B-1} c_b \frac{\partial  \langle \hat{P}_b \rangle}{\partial \theta_i}\nonumber \\
    &=& \sum_{b=0}^{B-1} \frac{c_b}{2} [ \hat{P}_b(\theta_i + \pi/2) \nonumber \\ & \quad& \quad - \hat{P}_b(\theta_i - \pi/2) ]
\end{eqnarray}
\noindent and as per Sect. \ref{subsec:richardson} they have variance
\begin{equation}
    (\sigma^2_{H_1})^\prime_i \approx \sum_{b=0}^{B-1} \frac{c_b^2}{2} (\sigma^2_{P_b})_{i+}, 
\end{equation}
\noindent where the notation $(\sigma^2_{P_b})_{i+}$ indicates the variance of the estimate of $\hat{P}_b$ when computing the shifted expectation value for the gradient of the $i$'th variational parameter. The approximation comes from the assumption that the variances at both of the shifted values are comparable \cite{mari2021estimating}. When ZNE with RE at $n$ different scale factors is used, the variance is amplified to
\begin{equation}
 (\sigma_{H_1}^2)_i^\prime = \sum_{b=0}^{B-1} \frac{c_b^2 }{2} \sum_{j=0}^{n}  \gamma_j^2  (\sigma^2_{P_b})_{ij+},
\end{equation}
\noindent where the new subscript $j$ on the variances indicates the scale factor.

The Hessian can also be computed using parameter-shift rules. In PennyLane, the rules derived in \cite{mari2021estimating} are used. They differ between the diagonal and off-diagonal elements, and consist of two and four terms, respectively: 
\begin{eqnarray}
\frac{\partial^2 \widehat{\langle H \rangle}}{\partial \theta_i^2} &=&\sum_{k=0}^{K-1} \frac{c_k}{2} [ \hat{P}_k(\theta_i + \pi) - \hat{P}_k(\theta_i) ] \\
\frac{\partial^{2} \widehat{\langle H \rangle}}{\partial \theta_i \partial \theta_\ell} &=& \sum_{k=0}^{K-1} \frac{c_k}{4}  [ \hat{P}_k (\theta_i + \pi/2, \theta_\ell + \pi / 2) \nonumber \\ && -  \hat{P}_k (\theta_i - \pi/2, \theta_\ell + \pi / 2) \nonumber \\   && - \hat{P}_k (\theta_i + \pi/2, \theta_\ell - \pi / 2)\nonumber \\
  && + \hat{P}_k (\theta_i - \pi/2, \theta_\ell - \pi / 2) ]
\end{eqnarray}
\noindent where it is implicit that all other $\theta_n$ are held constant during the shifting process. Consequently, their variances after error mitigation is applied are
\begin{eqnarray}
 (\sigma_{H}^2)_{ii}^{\prime\prime} &\approx& \sum_{k=0}^{K-1} \frac{c_k^2 }{2} \sum_{j=0}^{n}  \gamma_j^2  (\sigma^2_{P_k})_{iij},\\ 
 (\sigma_{H}^2)_{i\ell}^{\prime\prime} &\approx& \sum_{k=0}^{K-1} \frac{c_k^2 }{4} \sum_{j=0}^{n}  \gamma_j^2  (\sigma^2_{P_k})_{i\ell j+}.
\end{eqnarray}
\noindent where again the assumption is made that the variances at all shifted values are comparable (note that for the diagonal term, as the shifts are 0 and $\pi$, we do not indicate any shift in the subscript as we can simply consider the variance at the unshifted value). 

While the increase in variance due to mitigation is inevitable, it is possible to improve the accuracy of an estimate (i.e., reduce its standard error) by increasing the number of shots. Consider, for simplicity, the case of a single Pauli expectation value $\langle P \rangle$ computed using $N$ shots on a noisy device. If the variance of the estimate is $\sigma^2$, then the standard error of the mean is 
\begin{equation}
    \hbox{SEM} = \frac{\sigma}{\sqrt{N}}.
\end{equation}
\noindent Suppose error mitigation with $n$ scale factors is performed and each estimate is done using $M$ shots. The mitigated value, obtained according to Eq. \ref{eq:richardson_extrap} is $\hat{\mathcal{P}} = \sum_{j=0}^{n} \hat{P}_j \gamma_j$, and its standard error is 
\begin{equation}
    \hbox{SEM}_{mit} = \sqrt{\sum_{j=0}^n SEM^2_j} = \sqrt{\sum_{j=0}^n \frac{\gamma_j^2 \sigma^2_j}{M}}
\end{equation}
\noindent where here $\sigma^2_j$ indicates the variance of the estimate taken at the $j'$th scale factor. One can recover an expression for the minimum $M$ required to ensure the standard error does not increase by equating SEM and $\hbox{SEM}_{mit}$:
\begin{equation}
   M \geq \frac{N}{\sigma^2} \sum_{j=0}^n \gamma_j^2 \sigma^2_j.
\end{equation}
\noindent Clearly the required number of shots increases with the number of scale factors, and may grow quite large due to the factor of $\gamma_j^2$. As an illustrative example, the $4$-spin system with unitary folding at scale factors $1, 3, 5, 7$ was considered. A simple calculation was done on a single expectation value (the first shifted term of the gradient of $Z_0 Z_1$, $X_0 X_1$, and $Y_0 Y_1$ with respect to the first parameter, at the ground state parameters for $r=0.9$) An initial estimate was performed using $N=8192$ shots at each scale factor. The variances increase roughly linearly with scale factor, and the values of $\gamma_j^2$ are $\approx 4.79, \enskip 4.79, \enskip 1.72, \enskip 0.098$. The estimated number of shots is $M \approx 100000$, an increase of about $10$x. 

Thus, while the estimates can be improved, it is no surprise that this comes at a cost. However, this cost is further compounded by the need to evaluate expectation values at two or four parameter values in order to compute gradients and Hessians. While compensating for this in a simulation is as simple as increasing the number of shots, hardware configuration may lead to this being impractical. For instance, at the time of writing, the Manila device used for noisy simulation allows a maximum of 20000 shots per circuit, so multiple jobs would need to be submitted. Furthermore, since the Hamiltonians in question have more than one expectation value to compute, and the variances depend also on the Pauli coefficients, the compounding of additional circuits required makes obtaining a reasonable estimate on hardware impractical without dedicated access.\\
 
\section*{Appendix C: Additional results}
\label{appendix:more_results}

In this section additional numerical results are presented for the 4- and 6-spin systems. In particular, we show results for mitigated estimation of the ground state energy of both systems, and the absolute mitigation error \cite{qermit} (which, for the unmitigated case, is simply the error).

Note that error mitigation has the best overall performance for the estimation of the ground state energy; this is not surprising given that there is no dependence on the results on the solution of a noisy linear system. As 100 trials were performed at each $r$, there are two errors to plot: the absolute mitigation error for each trial, averaged over all trials; or the absolute mitigation error of the average of all results together. For the ground state energy of the $L=4$ system (Fig. \ref{fig:l4_gse_error}) one sees that a small amount of mitigation generally improves on the results of any individual trial. When considering all the mitigation results together, all mitigation methods see improvement. As each trial is independent, this provides insight into what could be expected in the limit of an infinite number of shots (in this case, obtained from running 819200 shots for each term in the Hamiltonian, for each circuit). For $L=6$ the benefits are more clear. This is not unexpected: given the small size of the 4-spin circuit, the increase in statistical noise that occurs when mitigating is worse than simply using the results as-is given the noise levels in today's processors. On the other hand for the 6-spin case, where there are five CNOTs in the circuit, mitigation is helpful.

\begin{figure*}[ht]
    \centering
    \includegraphics[width=\textwidth]{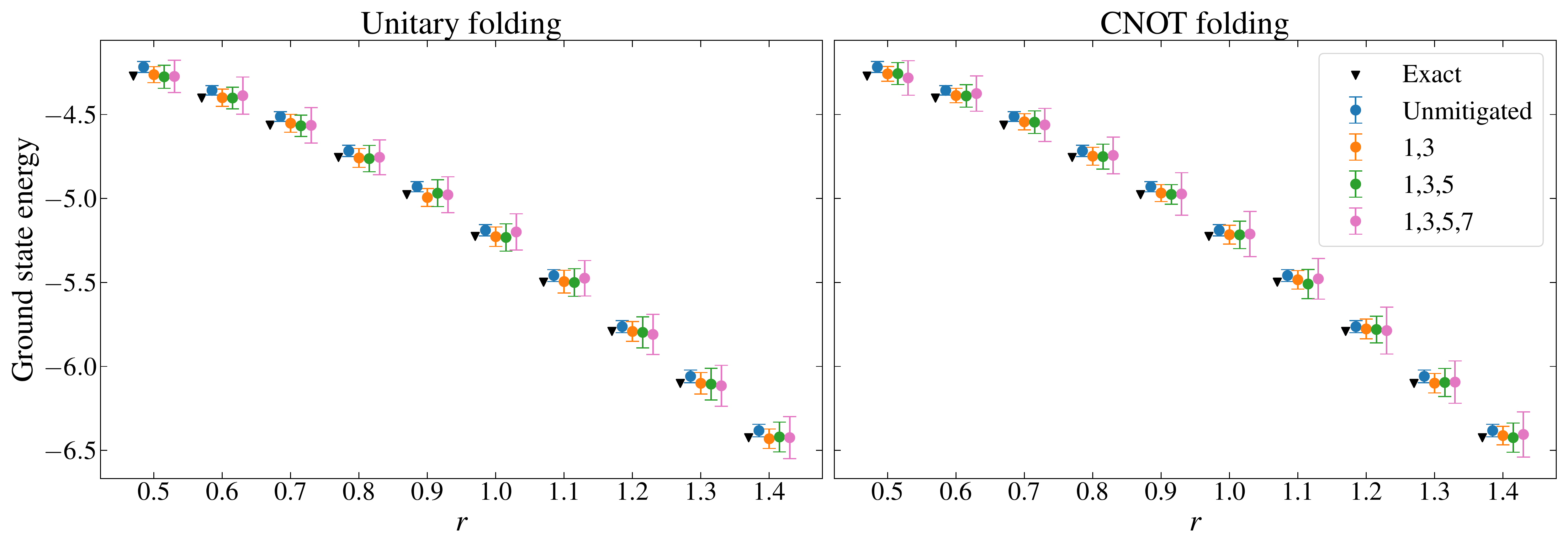}
    \caption{Error-mitigated ground state energies of the 4-spin system. Mean and standard deviation of 100 independent trials are shown for each value of $r$.}
    \label{fig:l4_gse}
\end{figure*}

\begin{figure*}[ht]
    \centering
    \includegraphics[width=\textwidth]{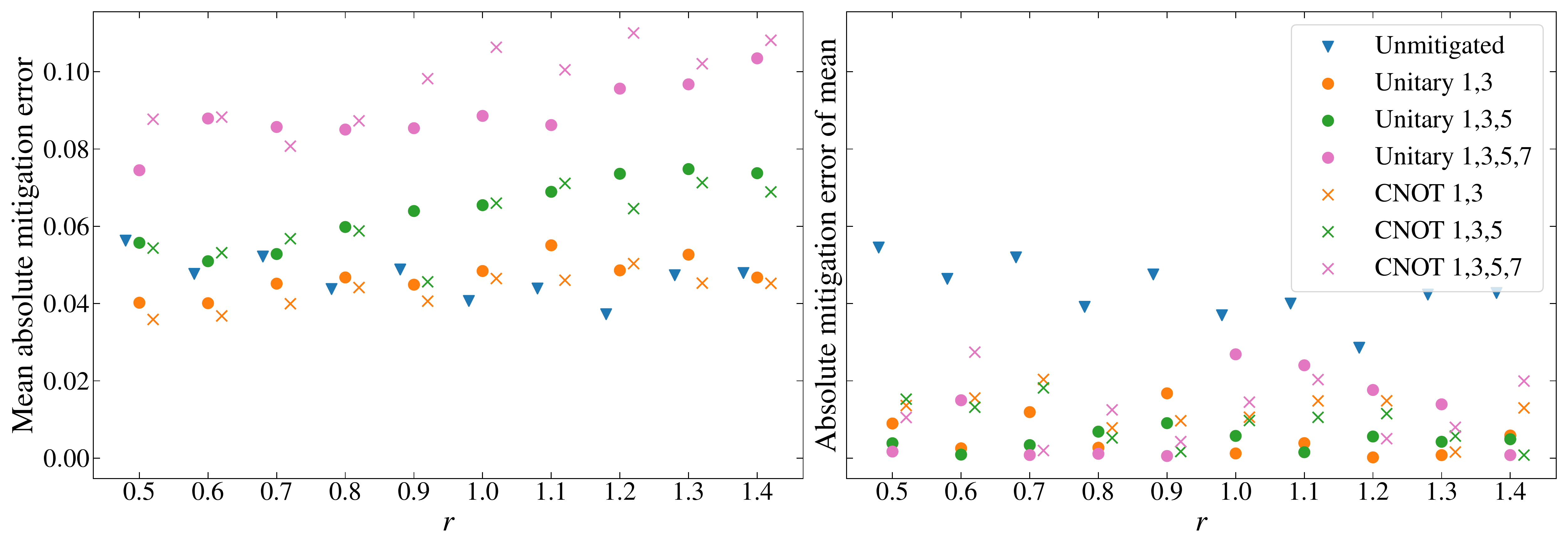}
    \caption{Mitigation error for the ground state energy of the 4-spin system. }
    \label{fig:l4_gse_error}
\end{figure*}

\begin{figure*}[ht]
    \centering
    \includegraphics[width=\textwidth]{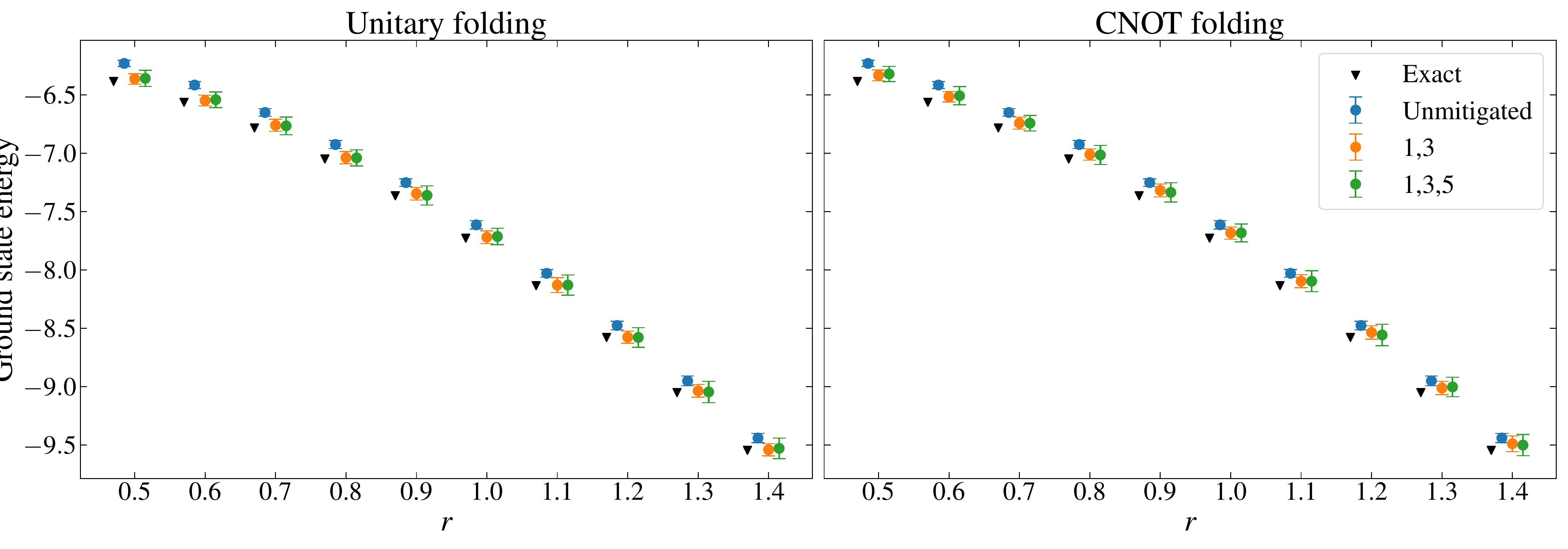}
    \caption{Error-mitigated ground state energies of the 6-spin system. Mean and standard deviation of 100 independent trials are shown for each value of $r$.}
    \label{fig:l6_gse}
\end{figure*}

\begin{figure*}[ht]
    \centering
    \includegraphics[width=\textwidth]{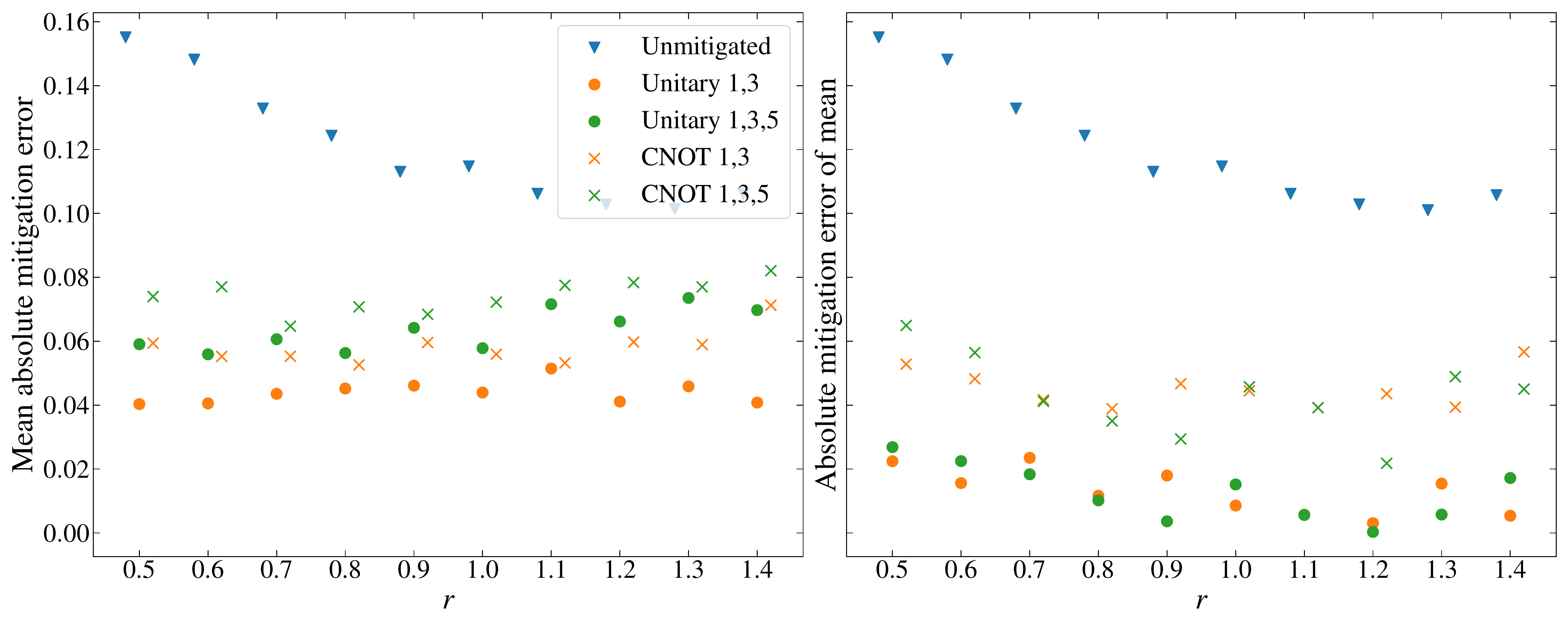}
    \caption{Mitigation error for the ground state energy of the 6-spin system. }
    \label{fig:l6_gse_error}
\end{figure*}

\begin{figure*}[ht]
    \centering
    \includegraphics[width=\textwidth]{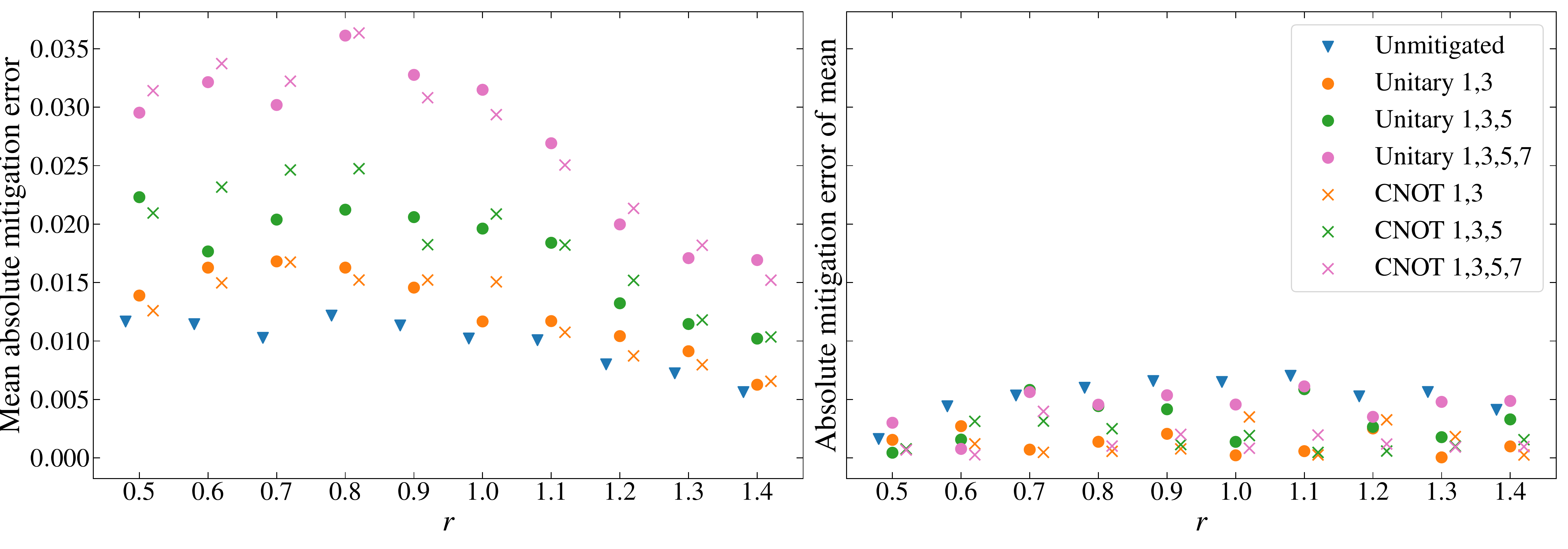}
    \caption{Mitigation error for the second energy derivative of the 4-spin system. }
    \label{fig:l4_d2e_error}
\end{figure*}

\begin{figure*}[ht]
    \centering
    \includegraphics[width=\textwidth]{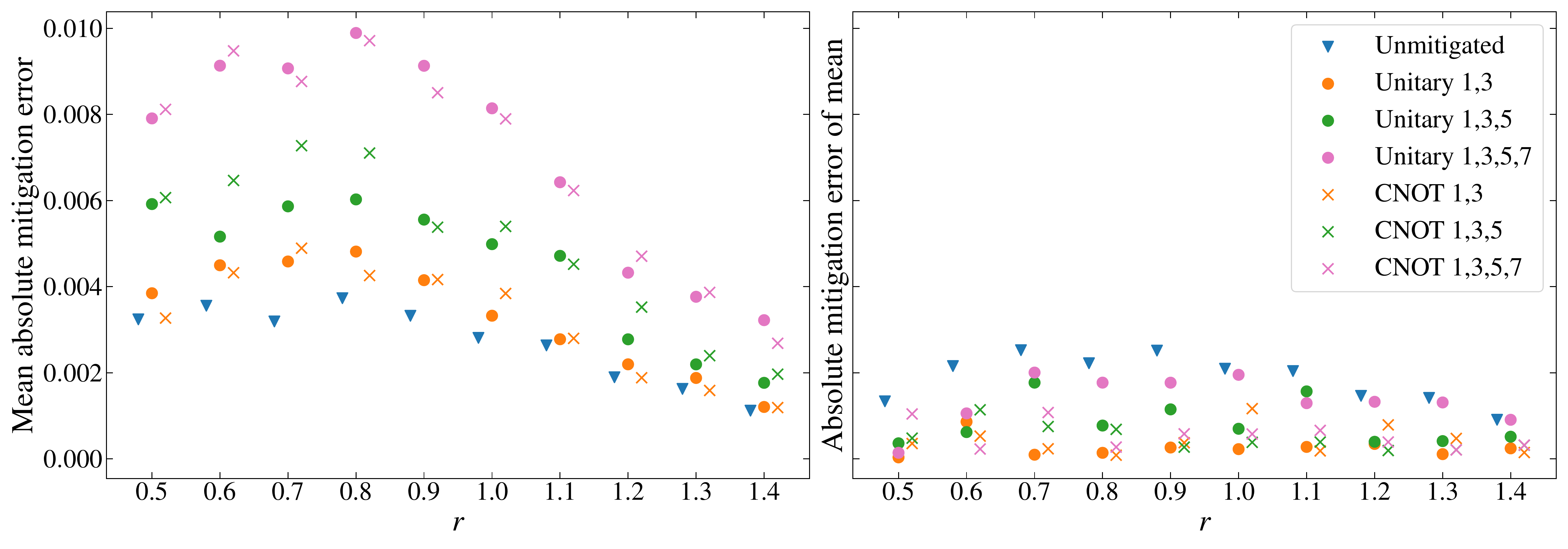}
    \caption{Mitigation error for the fidelity susceptibility of the 4-spin system. }
    \label{fig:l4_fs_error}
\end{figure*}

For the second energy derivative (Fig. \ref{fig:l4_d2e_error}) and fidelity susceptibility (Fig. \ref{fig:l4_fs_error} in the $L=4$ case), the benefits of mitigation in an average trial are less clear, but again in the limit of an infinite number of shots one can expect an advantage, at least up to a maximum of 3 folds; running with 4 folds produces worse results for both metrics, even for the 4-spin system (this data is provided in the GitHub repository). Running with two different scale factors and then extrapolating back produces the best results for both system sizes; while for the 4-spin case there is again no real improvement noted, there is a clear improvement in the $L=6$ case.

\begin{figure*}[ht]
    \centering
    \includegraphics[width=\textwidth]{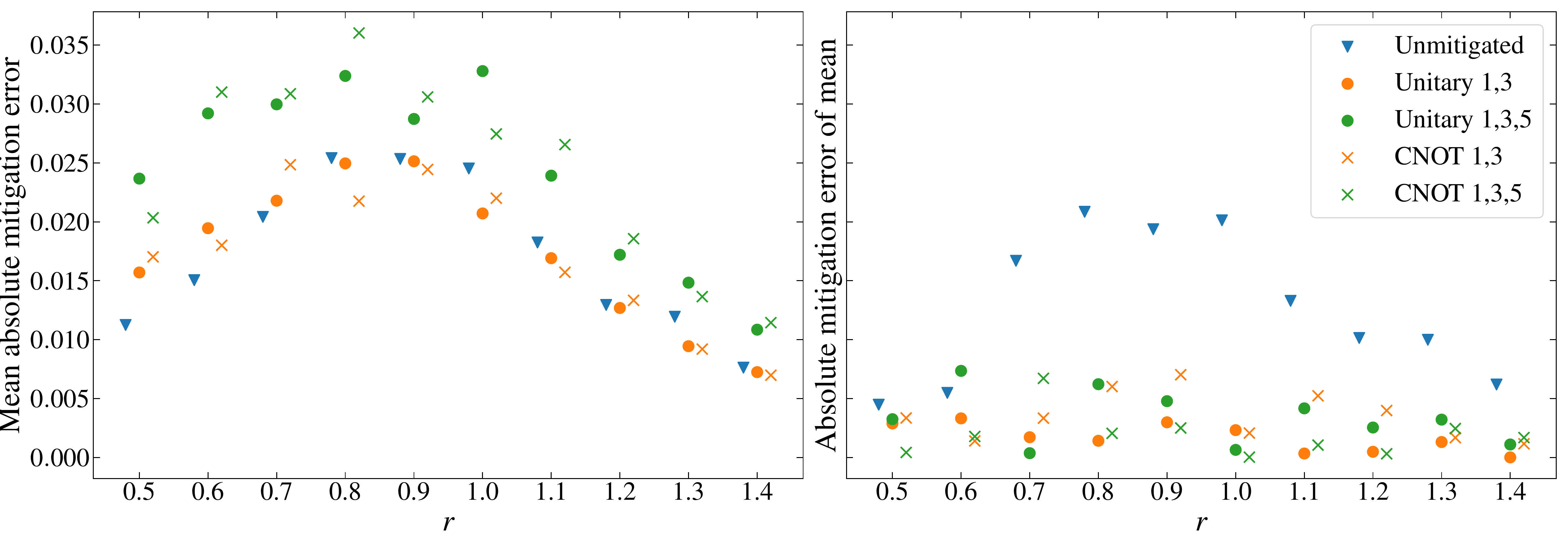}
    \caption{Mitigation error for the second energy derivative of the 6-spin system. }
    \label{fig:l6_d2e_error}
\end{figure*}

\begin{figure*}[ht]
    \centering
    \includegraphics[width=\textwidth]{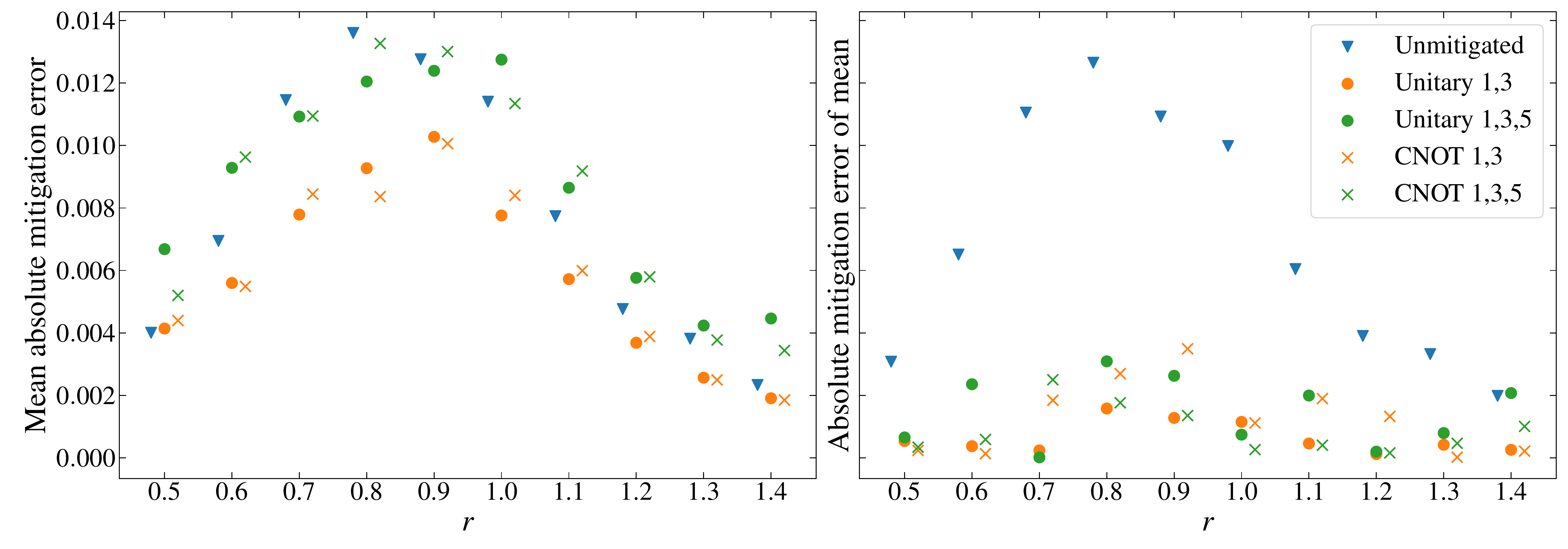}
    \caption{Mitigation error for the fidelity susceptibility of the 6-spin system. }
    \label{fig:l6_fs_error}
\end{figure*}

\bibliographystyle{h-physrev4}
\bibliography{main}

\end{document}